\documentclass[preprint,review,10pt]{elsarticle}

\usepackage{amssymb}
\usepackage{amsmath}

\usepackage{graphicx} 
\usepackage{amsmath}
\usepackage{amsthm}
\usepackage{graphicx}
\usepackage{subcaption}
\usepackage{color}
\usepackage{mwe}
\usepackage{hyperref}

\usepackage{ltablex,array}
\usepackage{longtable}
\usepackage{amssymb}
\usepackage{siunitx}
\usepackage{amsmath}
\usepackage{arydshln}
\usepackage{multirow}
\usepackage{booktabs}
\usepackage{hyperref}
\usepackage{graphicx} 
\usepackage{nameref}
\usepackage{wrapfig}
\usepackage{tabularx}
\usepackage[ruled,vlined,linesnumbered]{algorithm2e}
\usepackage{xcolor}
\usepackage{tcolorbox} 
\tcbuselibrary{skins} 
\usepackage{mathtools}
\usepackage{adjustbox}
\usepackage{graphicx}
\usepackage{booktabs}

\newtheorem{thm}{Theorem}
\newtheorem{lemma}{Lemma}
\newtheorem{property}{Property}
\newtheorem{corollary}{Corollary}
\newtheorem{definition}{Definition}

\newcommand\tasksmachine[1]{\mathcal{J}_{#1}}

\newcommand\first[2]{t_{#1}^{#2}}

\newcommand\start[2]{s_{#1}^{#2}}

\newcommand\larg{w}
\newcommand\hyperp{H}

\newcommand\rect[2]{R_{#1}^{#2}}
\newcommand\height[2]{h_{#1}^{#2}}
\newcommand\base[1]{b_{#1}}
\newcommand\job[1]{\mathcal{C}^{#1}}
\newcommand\jobperset[1]{{C}^{#1}}
\newcommand\task[2]{J_{#1}^{#2}}

\newcommand\proctime[2]{p_{#1}^{#2}}
\newcommand\per[1]{\tau(#1)}
\newcommand\perjob[1]{T^{#1}}

\newcommand\utilization[1]{U_{#1}}
\newcommand\utilinst{U}
\newcommand\orderTasks{J}
\newcommand\SE[1]{S^{#1}}
\newcommand\DGk[1]{D^{#1}}
\newcommand\DGkalpha[2]{D^{#1}(#2)}
\newcommand\DGm{D_{\max}}
\newcommand\DGmalpha[1]{D_{\max}({#1})}
\newcommand\DGs{D_{sum}}
\newcommand\DGsalpha[1]{D_{sum}({#1})}

\usepackage[capitalize]{cleveref}
\journal{Computers \& Operations Research}

\begin{document}

\begin{frontmatter}



\title{Periodic Chains Scheduling on Dedicated Resources - \newline A Crucial Problem in Time-Sensitive Networks}


\cortext[corr1]{Corresponding author}

\affiliation[dcefel]{organization={DCE, FEE, Czech Technical University in Prague}, addressline={Technická 2}, city={Praha}, postcode={160 00}, country={Czech Republic}}
\affiliation[ciirc]{organization={IID, CIIRC, Czech Technical University in Prague}, addressline={Jugoslávských partyzánů 1580/3}, city={Praha}, postcode={160 00}, country={Czech Republic},}
\affiliation[sorb]{organization={Sorbonne University, CNRS, LIP6}, addressline={F-75005}, city={Paris}, country={France}}
\affiliation[nat]{organization={University Paris Nanterre}, addressline={F-92000}, city={Nanterre}, country={France}}

\author[dcefel,ciirc]{Josef Grus\corref{corr1}}
\ead{josef.grus@cvut.cz}
\author[sorb,nat]{Claire Hanen}
\ead{claire.hanen@lip6.fr}
\author[ciirc]{Zdeněk Hanzálek}
\ead{zdenek.hanzalek@cvut.cz}

\begin{abstract}
Periodic messages transfer data from sensors to actuators in cars, planes, and complex production machines. When considering a given routing, the unicast message starts at its source and goes over several dedicated resources to reach its destination. Such unicast message can be represented as a chain of point-to-point communications. Thus, the scheduling of the periodic chains is a principal problem in time-triggered Ethernet, like IEEE 802.1Qbv Time-Sensitive Networks. This paper studies a strongly NP-hard periodic scheduling problem with harmonic periods, task chains, and dedicated resources. We analyze the problem on several levels and provide proofs of complexity and approximation algorithms for several special cases. We describe a solution methodology to find a feasible schedule that minimizes the chains' degeneracy related to start-to-end latency normalized in the number of periods. We use the local search with the first fit scheduling heuristic, which we warm-start with a constraint programming model. This notably improves the schedulability of instances with up to 100\% utilization and thousands (and more) of tasks, with high-quality solutions found in minutes. An efficient constraint programming matheuristic significantly reduces the degeneracy of the found schedules even further. The method is evaluated on sets of industrial-, avionic-, and automotive-inspired instances.

\end{abstract}


\begin{highlights}
\item Offset schedules guarantee worst-case degeneracy for acyclic instances
\item Constraint programming successfully initializes the local search
\item Per-period matheuristic efficiently improves existing schedules
\item Publicly available diverse set of PSP instances
\end{highlights}


\end{frontmatter}



\section{Introduction}
\label{sec:introduction}
\subsection{Motivation}
Periodic events and processes play an essential role in many application domains. To ensure efficient and deterministic data exchange among different parts of the application, engineers need to configure communication protocols to satisfy timing and throughput requirements. 
For example, with the advent of self-driving cars \cite{hanzalek2023} requiring a deterministic exchange of high-volume data, the communication network may quickly become a congestion point, and efficient schedules are needed to satisfy all safety-relevant features. When considering a given routing, the periodic (unicast) message goes from a sensor over several dedicated resources to reach its actuator. Thus, the message can be represented as a chain of point-to-point communications performed on communication links.

Ethernet is widely used in everyday life due to its high throughput and availability of SW stacks. Still, it lacks time determinism, essential for real-time applications, such as in-vehicle networks, industrial production lines, and audio streaming, that are very sensitive to delays. Time-sensitive networking (TSN) is a set of new standards that enables sending messages in a time-triggered manner. To use TSN efficiently and satisfy various requirements, many researchers  \cite{Stuber2023,Xue2024} have developed algorithms that can synthesize a schedule of periodic messages. Their results are versatile, but the network utilization and algorithm performance are often very low. Thus, in this paper, we omit a complex behavior of the TSN (link delays, queues, deadlines, frame isolation, see \cite{vlk2022largescale}), and we concentrate on a crucial part of this problem: the scheduling of periodic tasks. We focus on harmonic periods of the messages (i.e., for each pair of distinct periods, it holds that the longer period is divisible by the shorter one) that are widely used in engineering practice and help with the tractability of Periodic Scheduling Problem (PSP).

\subsection{Related Work}
Scheduling of periodic tasks has been studied from many different directions, as is outlined in \cite{minaeva2022survey}. Crucial results regarding the complexity of the scheduling of periodic tasks on a single resource were presented in \cite{cai1996nonpreemptive}, where the authors proved that the problem with non-preemptive tasks is strongly NP-complete. However, in \cite{korst1996scheduling}, it was shown that if both the periods and processing times of the tasks are harmonic, a feasible schedule can be found in polynomial time. 

Scheduling of both harmonic and non-harmonic periodic problems and several special cases were studied in \cite{marouf2010schedulability,marouf2011scheduling}. The authors presented a local schedulability condition for the general case of PSP.  
Motivated by the desire to increase the schedulability for highly-utilized instances, alternative views of PSPs based on bin-trees \cite{Eisenbrand2010} and bin packing \cite{lukasiewycz2009flexray,hanen2020periodic,grus2024icores} were presented. These perspectives enabled the relevant authors to devise heuristics and exact methods to find feasible schedules efficiently. 

PSPs are often found in the domains of avionics and automotive. Building on their previous work, authors of \cite{eisenbrand2010solving} used Integer Linear Programming (ILP) to partition tasks on different processors of modern aircraft. In \cite{AlSheikh2012avionicPeriodic}, the goal was to schedule tasks in the avionics domain to maximize a leeway between them. This way, the occasional increase in the processing time of the task would not interfere with the rest of the schedule. A game-theoretic approach to tackling a similar problem was presented in \cite{Pira2016gametheorertic}. In \cite{deroche2017avionicGreedyWithEnumearation}, the authors tackled the scheduling of so-called partitions on shared processors. Their idea utilized a greedy heuristic, streamlining an otherwise exhaustive approach without a significant loss in the quality of the schedules. In \cite{Blikstad2018relaxedsubaero}, communication between modules of the aircraft is considered, and the authors solve it with constraint-generating ILP procedure.

In the automotive domain, many papers tackle scheduling problems related to specific protocols and architectures. In \cite{monot2012autosarLptAndPartitioning}, the authors schedule tasks for AUTOSAR architecture to eliminate load peaks. ILP and heuristics for the same architecture were also investigated in \cite{becker2016autosarIlp}. FlexRay, a protocol often coupled with the AUTOSAR specification, was studied in \cite{lukasiewycz2009flexray} and more recently in \cite{dvorak2014multi,dvorak2019}. In the latter, the authors consider multiple message variants so the same schedule can be repurposed for different vehicle models.

Apart from communications, PSPs apply to manufacturing and maintenance. In \cite{Liao2003CorPerMaintanenca,xu2008parallelmaint}, the scheduling problems with (almost) periodic maintenance windows are studied; there, the maintenance task is frequently repeated, while the manufacturing tasks are essentially aperiodic. This framework is helpful in practice since regularly inspecting the machines reduces the risk of failures. A practical application can be found in \cite{wu2014patient}, where the authors schedule patients' non-preemptive treatments around the medical machine's maintenance.

The rest of this section focuses on papers regarding PSP with dedicated resources, which is the main focus of this paper. Messages are transmitted through the network, and the tasks are assigned to resources beforehand, given the network's routing.

In \cite{roy2016multi}, the authors co-optimize the resource utilization as well as the control performance of the system that relies on signals from sensors passed through the network. In \cite{minaeva2021control}, large-neighborhood search is used to improve control-wise properties of the schedules.

Another domain is scheduling for cloud radio access networks. In \cite{guiraud2018,GuiraudS24}, the authors focus on periodic scheduling with a single period and uniform processing times of the messages. Especially in the latter, the no-buffering constraint complicates the problem. The authors provide a polynomial scheduling algorithm applicable, when the network load is sufficiently low, and a general greedy method used in other cases. 

Early work on offline scheduling for TTEthernet (an example of time-triggered ethernet) was presented in \cite{steiner2008ttethernet}. The authors studied several network topologies, including line topology, and used satisfiability modulo theories (SMT) solver in their proposed incremental workflow. Many papers emerged with the introduction of the TSN; see survey papers  \cite{Stuber2023,Xue2024}. The authors categorized several methods and performed experimental comparisons in the latter. 

In \cite{serna2018tsnwindows}, the authors developed an SMT model to shorten gate control lists controlling the schedules; their length is a limited resource in practice. Characteristics of the real industrial networks were exploited in \cite{hellmanns2020industrial} to develop a two-stage scheduling method; authors were able to isolate parts of the problem and solve them independently. The TSN formalism can be further extended to include, e.g., authentication, as is demonstrated in \cite{reusch2022tsnwithauth}. The cryptography keys are disclosed with a time delay (enabled by the TSN) so that the network nodes can authenticate themselves. 

In \cite{vlk2020enhancing,vlk2022largescale}, the authors focused on a complete set of TSN requirements and tight time windows and achieved only low utilization of communication resources. On the other hand, in \cite{hladik2020complexity}, a first-fit heuristic for a very similar problem was presented (dedicated resources with soft constraints instead of hard ones), and the authors achieved nearly 100\% utilization of the resource. This observation triggered the research presented in this paper, concentrating on the principal aspects of scheduling periodic chains on dedicated resources. 

\subsection{Outline and Contributions}

In this work, we primarily focus on the scheduling of messages, represented by chains of non-preemptive tasks. Each task is to be performed on a dedicated resource, and all tasks of the chain have the same period from a set of harmonic periods. We optimize the degeneracy, which is related to the message's start-to-end latency normalized in the number of periods. We build on a single-resource scheduling paper \cite{grus2024icores} and extend the heuristic outlined in \cite{hladik2020complexity}.

In \cref{sec:problem-description}, we formally describe the problem and its structural properties, including the connection between PSP with harmonic periods and bin-packing. In \cref{sec:complexity-analysis}, we discuss the complexity of several special cases of PSPs. In \cref{sec:methods}, we describe a heuristic algorithm and Constraint Programming (CP) approaches to warm-start the algorithm and fine-tune the final solution. In \cref{sec:experiments}, we discuss the experimental results, and in \cref{sec:conclusion}, the conclusions are drawn. The list of the most used symbols follows. Furthermore, instances used in our experiments are available \cite{our-data}.

The main contributions of our paper are:
\begin{itemize}
    \item Proofs of the complexity of several classes of PSPs
    \item  Guarantees of maximum degeneracy for 
    instances with acyclic support graphs and a bottleneck resource
    \item CP warm-start approach for highly-utilized instances
    \item CP-based matheuristic for improving existing schedule
\end{itemize}
\section{Periodic Scheduling with Dedicated Resources}
\label{sec:problem-description}
\subsection{Model Abstraction}

In this paper, we use concepts of tasks $\task{}{}$, chains of tasks $\job{}$ and resources (described in detail in \cref{sec:problem-statement}) which can be used to model diverse real-life systems and problems. In periodic manufacturing, the resources would correspond to production machines, chains to orders that need to be processed by several machines, and tasks to operations of the order.

Specific hardware and software of the communications network influence how the model of the associated PSP is created. For example, when the CPU of the network node performs time-consuming processing on each message, then the CPU itself is the resource since it is the communication bottleneck.
For example, in \cite{guiraudthese,GuiraudS24}, tasks correspond to computations in the network nodes, and passing a message through a link induces a precedence delay between the tasks.

In modern TSN applications, the network nodes use massive parallel hardware. Thus, the network node itself is not the communication bottleneck. Therefore, the resources correspond to communications links between network nodes. Chains correspond to messages sent from one network node to another, and the task is a point-to-point communication via a single link. We elaborate on the relationship between the network topology and associated PSP in further detail in \cref{fig:gant_instance} with four network nodes A, B, C, and D depicted by blue rectangles, and resources 1-6 shown as blue arrows with ellipses.

Our model and benchmarks also capture other realistic phenomena. For example, in automotive, zonal architecture is often used as a communication architecture, where the peripheral links connected to sensors have much slower speeds than the links connecting central nodes. This is modeled by proportionally increasing the processing times of tasks dedicated to slower links. 

\subsection{Problem Description}\label{sec:problem-statement}
\subsubsection{Tasks, Resources}
Consider a set of $\mu$ chains of tasks. Each chain corresponds to a message sent periodically. In the rest of the paper superscript $k$ is used to index the chain, while subscripts are used to describe the task number within the chain. So, a superscript does not indicate a power function.  
\begin{definition}[Tasks, chains, and resources] A chain $\job{k}$ with $k\in\{1,\ldots,\mu\}$ is composed of tasks $\task{i}{k}$, $i\in\{1,\ldots,n^k\}$ with integer processing times $\proctime{i}{k}$.  The tasks of a chain $\job{k}$ are periodic and share the same period $\perjob{k}$. Once a task $\task{i}{k}$ is started, it is repeated every $\perjob{k}$ time unit.
Each task and its repetitions are to be performed on a pre-defined unit capacity resource $\{1,\ldots,M\}$. 
 The index of the resource processing $\task{i}{k}$ is denoted by $\pi_i^k$. 
For a resource $m\in\{1,\ldots,M\}$ we denote by $\tasksmachine{m}$ the set of tasks $\task{i}{k}$ assigned to it (i.e. such that $\pi_i^k=m$).
\end{definition}

The structure of an instance will be analyzed through the notion of a support graph:
\begin{definition}[Support graph]
    We call the support graph of an instance the directed graph $G=(V,E)$ where $V=\{1,\ldots, M\}$ represents the resources, and there is an arc from node $m$ to node $m'$ if there is a chain $\job{k}$ such that two consecutive tasks of the chain $\task{i}{k},\task{i+1}{k}$ satisfy $\pi_i^k=m,\pi_{i+1}^k=m'$.
\end{definition}

\begin{figure}
        \centering
        \begin{subfigure}[b]{0.49\textwidth}  
            \centering 
            \includegraphics[width=\textwidth]{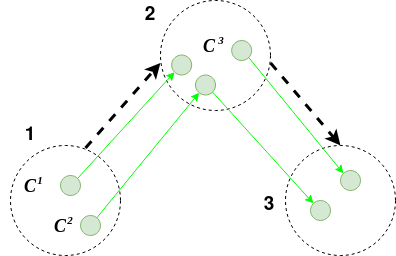}
            \caption{Support graph with three vertices and without directed cycles.}
            \label{fig:support-acyclic}
        \end{subfigure}
        \hfill
        \begin{subfigure}[b]{0.49\textwidth}
            \centering
            \includegraphics[width=\textwidth]{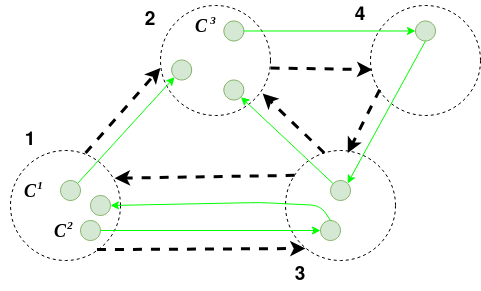}
            \caption{Support graph with four vertices and with directed cycles produced by chains $\job{2}, \job{3}$.}    
            \label{fig:support-cyclic}
        \end{subfigure}
        \caption{Examples of support graphs. Their vertices (i.e., resources) are the dashed ellipses, and their edges are shown as dashed arrows. The green chains and green tasks induced the edges of the support graphs.} 
        \label{fig:supports}
\end{figure}

Examples of such support graphs are shown in \cref{fig:supports}, including the chains which induced the graphs' edges.

\subsubsection{Harmonic Periods, Feasible Schedules}

In this paper, we assume that all periods belong to a harmonic set:

\begin{definition}[Harmonic set of periods] For any chain $\job{k}$, its period $\perjob{k}$ is in an harmonic set of integers $\{\per{1},\ldots,\per{\rho}\}$. Such a set  satisfies \begin{equation}\forall l\,\in\nolinebreak\{2,\ldots,\rho\}, \per{l}=\base{l}\per{l-1}\end{equation} where $\base{l}$ is an integer.  The least period is also denoted $\larg=\per{1}$, and we define the  hyperperiod \begin{equation}
    \hat\hyperp=lcm(\per{1},\ldots,\per{\rho})=\per{\rho}
\end{equation}. \end{definition}

 We assume that all of the tasks of chain $\job{k}$ have the same period $\perjob{k}=\per{\beta^k}$, where $\beta^k$ is the index of period $\perjob{k}$ in the ordered set of periods.

\begin{definition}[Schedule]
A schedule assigns an integer start time $\first{i}{k}$ to the first occurrence of each task $\task{i}{k}$. Moreover, an occurrence of $ \task{i}{k}$ is performed at each time $\first{i}{k}+\alpha\perjob{k}$, for any integer $\alpha\ge 0$.
\end{definition}

A schedule is feasible if it satisfies precedence constraints and resource constraints  defined below:

\begin{definition}[Precedence constraints] Task $\task{i}{k}$ precedes task $\task{i+1}{k}$:
\begin{equation}\first{i+1}{k}\ge \first{i}{k}+\proctime{i}{k}\quad \forall k\in \{1,\ldots,\mu\}, \forall i\in\{1,\ldots,n^k-1\},\quad\label{eq:prec}\end{equation}
\end{definition}

\begin{definition}[Resource constraints] At any time instant $s$, a resource $m$ performs at most one task:
\begin{multline}\left\vert\left\{\task{i}{k} \in\tasksmachine{m} ,\quad\exists l\in \mathbb{N},\quad  \first{i}{k}+l\cdot\perjob{k}\le s<\first{i}{k}+\proctime{i}{k}+l\cdot\perjob{k}\right\}\right\vert\le 1\\\forall s\ge 0, \forall m\in \{1,\ldots,M\},\label{eq:machine}\end{multline}
\end{definition}

\begin{definition}[Utilization] We define the utilization of resource $m\in\{1,\ldots,M\}$ as:
\begin{equation}\utilization{m}=\sum_{\task{i}{k}\in\tasksmachine{m}}\frac {\proctime{i}{k}}{\perjob{k}}\end{equation}
The utilization of an instance is the maximum utilization of its resources.
\end{definition}

A necessary condition of the feasibility of a set of periodic tasks on a resource $m$ is that the utilization, i.e., the percentage of time resource $m$ is busy, should not exceed $1$. The fact that utilization is not greater than 1 implies that for each task $\proctime{i}{k}<\perjob{k}$ (otherwise the task $\task{i}{k}$ would be performed without any other task on its machine).

\subsubsection{Objectives}

All objective functions we consider are related to the start-to-end latency of chains, that is, the delay between the moment when the chain's first task started and the chain's last task ended. In communications, this refers to the delay between the moment the message was sent from the first (source) network node and is being processed by the first link (resource) and when the message was delivered via the last link to the last (sink) network node. However, minimizing the start-to-end latency of chains might lead to unnecessarily difficult problems. Previous work \cite{hanen2012grouping} has shown that measuring the number of periods crossed by a message is significant enough in practice and simplifies the problem's complexity. So, we consider period-normalized objectives as in \cite{hladik2020complexity}. Moreover,  discrete-time control of cyber-physical systems naturally matches such objectives. The controller operates with its sampling period \cite{ogata-1987a}, and if the message arrives even slightly after the end of the sampling period, it will need to wait until the end of the next sampling period. There is no gain in arriving earlier within the sampling period.

Let $t$ be a schedule. We define the following functions:
\begin{definition}
[Start-to-end latency of chain $k$] \begin{equation} \SE{k}\coloneqq\first{n^k}{k}+\proctime{n^k}{k} -\first{1}{k}\end{equation}
\end{definition}
\begin{definition}[Degeneracy of chain $k$] \begin{equation} \DGk{k}\coloneqq \left\lceil \frac{\SE{k}}{\perjob{k}}\right\rceil-1.\end{equation}
\end{definition} The degeneracy is the start-to-end latency normalized in the number of periods minus 1. Thus, a chain with start-to-end latency shorter than its period has degeneracy zero.
\begin{definition}[$\alpha$-degeneracy of a chain $k$] Given $\alpha\in (0,1]$: \begin{equation} \DGkalpha{k}{\alpha}\coloneqq \left\lceil \frac{\SE{k}}{\alpha\cdot \perjob{k}}\right\rceil-1.\end{equation} \end{definition}This function uses finer granularity of the degeneracy whose minimization might be more restrictive: for example, $\DGkalpha{k}{0.5}=0$, which means that the start-to-end latency of chain $\job{k}$ fits in half of the chain period.

For each measure, we consider the maximum value and the sum of values as objective functions to be minimized: 

    \begin{definition}[Objective functions]
    We call maximum degeneracy the objective\begin{equation} \DGm\coloneqq\max_{k=1,\ldots,\mu}\DGk{k}\end{equation} We call degeneracy the sum of degeneracies of all chains:\begin{equation} \DGs\coloneqq\sum_{k=1}^\mu \DGk{k}\end{equation} 
   Similarly  can be defined the maximum and sum $\alpha$ degeneracy $\DGmalpha{\alpha},\DGsalpha{\alpha}$.
 
   
   \end{definition}

\sloppy
According to the 
standard Graham notation extended to handle PSPs in \cite{minaeva2022survey}, our problem is denoted by ${PD|T_i^{harm},jit=0, chains|objective}$. $PD$ stands for parallel dedicated processors, $ chains$ for chains of tasks, $T_i^{harm}$ for harmonic periods, $jit=0$ since no jitter is allowed (i.e., the starting time of a task is entirely determined by the starting time of its first occurrence and by its period). We recall that all tasks of a chain have the same period.
\fussy

\subsection{Special Properties}
In the paper, we consider both general $PD|T_i^{harm},jit=0,chains|objectives$ instances, as well as instances generated with a set of realistic properties. For example, we generate instances with specifically engineered chains and assignment of tasks to resources, so the instance models periodic communications over a network with a given topology. Some of these properties, which will be used in the following sections of the paper, are:

\sloppy
\begin{description}
\item[Limited number of resources: ] This is described in the first field of Graham notation. For example, $PD2$  indicates two resources.
\item [Harmonic processing times: ] In this case, all values of processing times and periods are together in a harmonic set. This makes the single resource feasibility problem polynomial \cite{korst1996scheduling}. Graham notation for such case is $PD|(T_i,p_i)^{harm},jit=0, chains|objectives$.

\item [Topology:]  In this case, the underlying network topology is organized in such a way that the support graph of the instance has some special structure (e.g., a line or a tree). 

\item[Equal processing time on a chain: ] In this case, inspired by the most common application of communication protocols,   
point-to-point communications have the same duration since the packets have the same length and communication links have the same speed. In such case  $\proctime{i}{k}=\proctime{}{k}$. 

\item[Acyclic: ]
In this setting, we assume the support graph contains no directed cycle. Proofs related to this case are described in \cref{sec:complexity-analysis}. 
\item[Bottleneck resource: ]
A special case where there is at least one resource $b$ such that all chains go through $b$.

\item[Scattering: ] A special case of an acyclic instance with a bottleneck resource $b$ where all the first tasks of the chains are assigned to $b$ (as in the context of parallel computing \cite{grama03}).  
\end{description}

\subsection{Structural Properties}\label{sec:structural-properties}
In this subsection, we recall some structural properties useful to understand our algorithms and proofs of the next sections. First, we describe how to build a feasible schedule from a core schedule that only satisfies the resource constraints. Then, we describe the analogy introduced in \cite{lukasiewycz2009flexray} and formally established in
 \cite{grus2024icores}, of a schedule on one resource and a 2D packing problem.
The packing view is a crucial inspiration behind the warm-start approach described later.
\subsubsection{Building a Schedule from Independent Resource Schedules}
Let us consider a resource $m$ and the subset of tasks $\tasksmachine{m}$ assigned to $m$. We first define the following concept:
\begin{definition}[Core schedule]
We call core schedule on resource $m$ any schedule $\sigma$ of the tasks $\tasksmachine{m}$ that satisfies the resource constraint, and such that \begin{equation}\forall\task{i}{k}\in \tasksmachine{m},\sigma_i^k<\perjob{k}\end{equation}
\end{definition}
Obviously, if $t$ is a schedule of tasks $\tasksmachine{m}$ that satisfies the resource constraints on $m$, then the following $\sigma$ is a core schedule on $m$.
\begin{equation}
\sigma_i^k=\first{i}{k}\mod\perjob{k}, \quad\forall\task{i}{k}\in \tasksmachine{m}
\end{equation}

Note that if $t$ is not integer, then the $\mod$ operation here is defined as follows:
$\first{i}{k}=q\cdot\perjob{k}+\sigma_i^k$, with an integer $q$, and $\sigma_i^k\in[0,\perjob{k})$.

We now prove that from a core schedule $\sigma$ on each resource, we can build a feasible schedule with core $\sigma$ that satisfies the resource and the precedence constraint and minimizes the degeneracy among all schedules with core $\sigma$. Indeed,  we can satisfy the precedence constraints \eqref{eq:prec} of each chain $k$  by postponing a subsequent task in the chain to the next period whenever needed to satisfy the precedence constraint. For this purpose, we define \cref{algobase}.

\begin{algorithm}
    \DontPrintSemicolon
\SetAlgoLined
\SetNoFillComment
\LinesNotNumbered 
 \SetKwInOut{Input}{Input}
\SetKwInOut{Output}{Output}
\caption{Building a feasible schedule from a core schedule}
\label{algobase}
\Input{Chains, tasks, resources, core schedule $\sigma$ on each resource}
\Output{Schedule $\first{i}{k}=\sigma_i^k+q_i^k\perjob{k}$, with $q_i^k\in \mathbb{N}$ for each task $\task{i}{k}$}
\For{$k\leftarrow 1$ \KwTo $\mu$}{
$q_1^k\gets 0$ \;

\For{$i\leftarrow 1$ \KwTo $n^k - 1$}{
   \begin{math}q_{i+1}^k\gets \left\{\begin{array}{lll}q_i^k&if&\sigma_i^k+\proctime{i}{k}\le\sigma_{i+1}^k\\
   q_i^k+1&if&\sigma_{i+1}^k<\sigma_i^k+\proctime{i}{k}\le\sigma_{i+1}^k+\perjob{k}\\q_i^k+2&&otherwise\end{array}\right.\end{math}\;
    }
}
\end{algorithm}

The following lemma establishes that the schedule built by  \cref{algobase} is indeed feasible with respect to the precedence constraints and has minimal degeneracy among schedules with core $\sigma$.
\begin{lemma} 
Assume that we are given for each $m\in \{1,\ldots,M\}$, a core schedule $\sigma$ of tasks in $\tasksmachine{m}$. \cref{algobase} outputs a schedule $t$  satisfying the precedence constraints and for which the degeneracy of each chain $\job{k}$ is minimal among all schedules with core schedule $\sigma$.
\label{postpone-lemma}
\end{lemma}
\begin{proof}
Notice first that any schedule $\hat{\first{}{}}$ with core $\sigma$ satisfies, for each chain $\job{k}$ and task $\task{i}{k}$:
\begin{equation}
\hat{t}_i^k=\sigma_i^k +\gamma_i^k\perjob{k}
\end{equation}
with $\gamma_i^k\in\mathbb{N}$, since $\hat{t}_i^k\mod\perjob{k}=\sigma_i^k$. So the schedule computed by \cref{algobase} where $\first{i}{k}=\sigma_i^k+q_i^k\perjob{k}$ has indeed core $\sigma$.

Consider now the start-to-end latency of chain $\job{k}$ for any schedule $\hat{\first{}{}}$ with core $\sigma$:  \begin{equation}\SE{k}(\hat{\first{}{}})=\hat{t}_{n^k}^k+\proctime{n^k}{k}-\hat{t}_1^k=\sigma_{n^k}^k+\proctime{n^k}{k}-\sigma_{1}^k + (\gamma_{n^k}^k-\gamma_1^k)\perjob{k}\end{equation} so: \begin{equation}\DGk{k}(\hat{\first{}{}})=\gamma_{n^k}^k-\gamma_1^k+\left\lceil\frac{\sigma_{n^k}^k+\proctime{n^k}{k}-\sigma_{1}^k}{\perjob{k}}\right\rceil-1\end{equation}

So, among schedules with core $\sigma$, the degeneracy just depends on the difference $\gamma_{n^k}^k-\gamma_1^k$.

We now prove that the schedule computed by \cref{algobase} satisfies the precedence constraint and that $q_{n^k}^k-q_1^k$ is minimal for any chain $\job{k}$.

Consider a task $\task{i}{k}$ and its subsequent task $\task{i+1}{k}$. Notice that the corresponding precedence constraint on any  schedule  with core $\sigma$ is expressed as follows:
 \begin{equation}\hat{t}_{i+1}^k\ge\hat{t}_i^k+\proctime{i}{k}\quad\Leftrightarrow\quad\gamma_{i+1}^k-\gamma_i^k\ge \left\lceil\frac{\sigma_i^k+\proctime{i}{k}-\sigma_{i+1}^k}{\perjob{k}}\right\rceil\label{eqprec}\end{equation}
Notice that \cref{algobase} distinguishes three cases. If $\sigma_i^k+\proctime{i}{k}\le\sigma_{i+1}^k$ then:

\begin{equation}
    q_{i+1}^k-q_i^k=0= \left\lceil\frac{\sigma_i^k+\proctime{i}{k}-\sigma_{i+1}^k}{\perjob{k}}\right\rceil,
\end{equation}
so the schedule $\first{}{}$ satisfies equation \eqref{eqprec}. 

Similarly, if $\sigma_{i+1}^k<\sigma_i^k+\proctime{i}{k}\le \sigma_{i+1}^k+\perjob{k}$, then: 
\begin{equation}
    q_{i+1}^k-q_i^k=1=\left\lceil\frac{\sigma_i^k+\proctime{i}{k}-\sigma_{i+1}^k}{\perjob{k}}\right\rceil,
\end{equation} 
and equation \eqref{eqprec} holds.

In the latter case, we necessarily have $\sigma_i^k+\proctime{i}{k}<\sigma_{i+1}^k+2\perjob{k}$,  since $\sigma_i^k$ and $\proctime{i}{k}$ are lower than $\perjob{k}$. Then: 
\begin{equation}
    q_{i+1}^k-q_i^k=2=\left\lceil\frac{\sigma_i^k+\proctime{i}{k}-\sigma_{i+1}^k}{\perjob{k}}\right\rceil,
\end{equation}
and equation \eqref{eqprec} holds.

Moreover in the three subcases, for any other schedule $\hat{\first{}{}}$ with core $\sigma$ we have \begin{equation}
    \gamma_{i+1}^k-\gamma_i^k\ge q_{i+1}^k-q_i^k
\end{equation}

Hence, by summing these inequalities:
\begin{equation}
    \gamma_{n^k}^k-\gamma_1^k\ge q_{n^k}^k-q_1^k.
\end{equation}
So, the degeneracy of each chain $\job{k}$ is minimal for the schedule build by \cref{algobase}.

\end{proof}

\begin{figure}
\centering
\begin{subfigure}[b]{0.99\textwidth}
\includegraphics[width=0.99\textwidth]{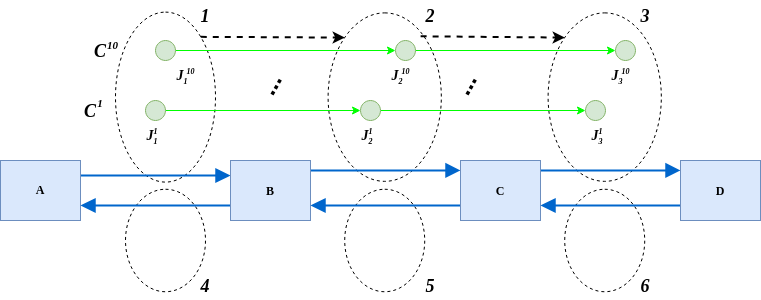}
   \caption{The network consists of four network nodes A, B, C, D in line topology connected with full-duplex links. Ten messages are sent from the leftmost network node A to the rightmost node D via three blue links (i.e., resources). Corresponding chains $\job{1}-\job{10}$, each consisting of three tasks $\task{i}{k}$, are to be scheduled on resources 1, 2, 3 (assignment of tasks to resources $\pi_i^k=i$ is shown by dotted ellipses). For simplicity, no messages are sent via resources 4, 5, 6 in the opposite direction.}
   \label{fig:gant_instance}
\end{subfigure}
\vskip\baselineskip
\begin{subfigure}[b]{0.99\textwidth}
   \includegraphics[width=0.99\textwidth]{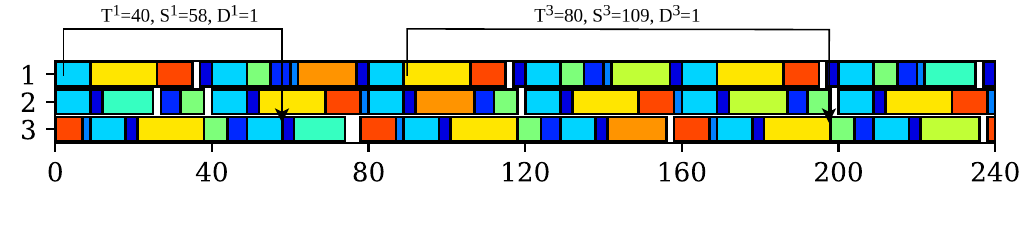}
   \caption{Schedule of ten chains on three resources spanning over one hyperperiod. Tasks of the same chain have the same color. Total degeneracy is 2. Start-to-end latency $\SE{k}$ is shown for light blue $\job{1}$ and yellow $\job{3}$ chains, which both have degeneracy $\DGk{k}=1$.}
   \label{fig:gant_schedule}
\end{subfigure}
   \caption{Illustrative communication case and associated schedules for instance with periods $\per{1}=40,\per{2}=80,\per{3}=240$. Ten chains are to be scheduled; $\job{1}$ and $\job{2}$ (dark and light blue) have period 40 (thus $\perjob{1}=\perjob{2}=\per{1}$, meaning $\beta^1=\beta^2=1$), $\job{3}-\job{7}$ (e.g., yellow) have period 80, and $\job{8}-\job{10}$ (e.g., cyan)  have period 240. Notice that all the messages are sent from the leftmost network node to the rightmost one; communication links in the other direction are not used. Utilization $\utilization{m}=0.96$ holds for all resources.}
   \label{fig:gant_example} 
\end{figure}

Such schedules, shown within a single hyperperiod, can be illustrated as in \cref{fig:gant_schedule}. There, 10 chains are delivered using a sequence of resources 1, 2, and 3, as is shown in \cref{fig:gant_instance}. Precedence constraints for chains hold, and no overlaps occur on either resource.  
\subsubsection{Single Resource Core Schedule Coordinates}\label{sec:shifted-core-schedule}
Consider a single resource $m$ and a feasible core schedule $\sigma$ of the independent periodic tasks of $\tasksmachine{m}$. Let $o_m$ be the start time of the first task $\task{f}{k}$ on resource $m$ with the least period $\larg$. We consider $o_m$ as the offset for resource $m$ from 0. So we define the shifted core schedule $\start{i}{k}$ for each task  $\task{i}{k}\in \tasksmachine{m}$ such that: $$\start{i}{k}=\left\{\begin{array}{ll}\sigma_{i}^{k}-o_m & if \ \sigma_{i}^{k}\ge o_m\\
\sigma_{i}^{k}+\perjob{k}-o_m&otherwise\end{array}\right.$$

 We call time interval $[o_m,o_m+\hat\hyperp)$ the ``observation interval''. Notice that for any integer $v\ge 1$, the schedule $\sigma$ of time interval $[o_m+(v-1)\hat\hyperp,o_m+v\hat\hyperp)$ on resource $m$ is the same as the schedule of the observation interval shifted by $(v-1)\hat\hyperp$ time units. This offset may differ for different resources. 

  \subsubsection{Associated HD2D Packing Problem} \label{sec:packing-section}
We can now define a height-divisible 2D (HD2D) packing instance associated with a periodic scheduling instance with harmonic periods.
Consider an instance $I$ of the scheduling problem. 

To each task $\task{i}{k}\in\tasksmachine{m}$ we associate a rectangle $\rect{i}{k}$ of width $\ell_i^{k}=\proctime{i}{k}$ and height $\height{i}{k}=\frac{\hat\hyperp}{\perjob{k}}$, to fit in a bin of width $\larg$ and height $\displaystyle \hyperp=\frac{\hat\hyperp}{\larg}$ without overlapping rectangles. Moreover, the packing has to be height divisible, which means that the vertical coordinate of each rectangle is an integer multiple of its height.
\begin{thm}[\cite{grus2024icores}]
There exists a feasible schedule of tasks of $\tasksmachine{m}$ on machine $m$ if and only if there exists a height divisible packing of the rectangles of the associated instance in a bin of width $\larg$ and height $\displaystyle \hyperp=\frac{\hat\hyperp}{\larg}$.
\end{thm}

\begin{figure}
\centering
\begin{subfigure}[b]{0.9\textwidth}
\includegraphics[width=0.99\textwidth]{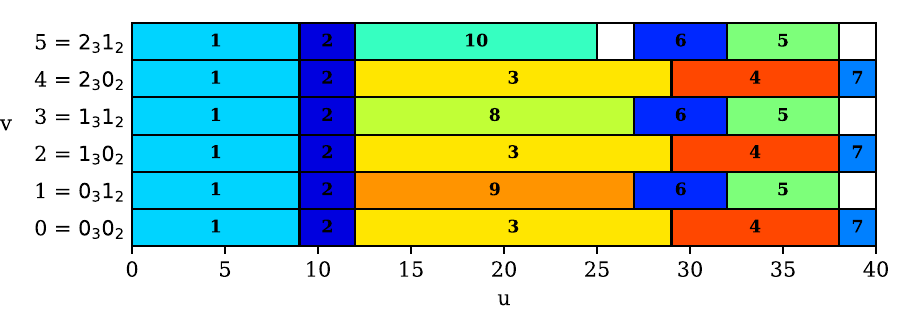}
   \caption{Schedule of a single observation interval of resource 2 arranged in rows of length 40.}
   \label{fig:sample_hyperperiod}
\end{subfigure}
\vskip\baselineskip
\begin{subfigure}[b]{0.9\textwidth}
   \includegraphics[width=0.99\textwidth]{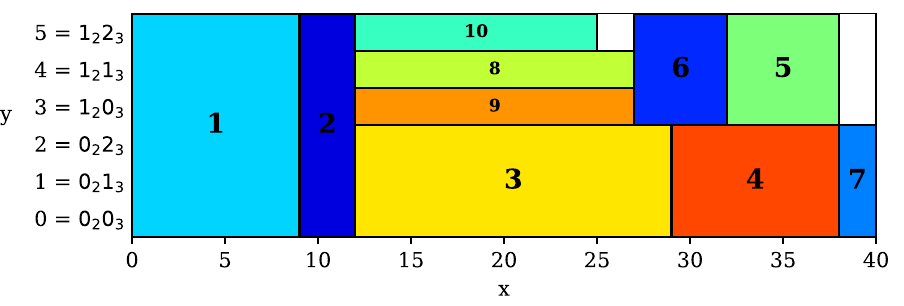}
   \caption{Height-divisible 2D packing of resource 2.}
   \label{fig:sample_placement}
\end{subfigure}

\caption{Observation interval and packing perspective of a schedule of resource 2 from \cref{fig:gant_schedule}. The offset of the shifted core schedule is $o_2=40$.}
\label{fig:sample_figures}
\end{figure}

Paper \cite{grus2024icores} gives the formal correspondence between the start time of a task in the core schedule $\start{i}{k}$ and coordinates of the bottom left corner of the associated rectangle. To give an intuition of this correspondence, we develop it on an example in \cref{fig:sample_figures}. The main idea is to arrange the periodic schedule in rows and shuffle them to produce the packing.

In \cref{fig:sample_hyperperiod}, the single ``observation interval'' associated with the schedule of resource 2 from the motivational example of \cref{fig:gant_schedule} is shown. Ten chains $\job{1},\dots,\job{10}$ pass through the resource. $\job{1}, \job{2}$ have the shortest period $\per{1} = 40$. Chains $\job{3} - \job{7}$ have a period equal to 80; the rest have a period equal to 240, which is the hyperperiod. Origin $0$ is fixed by the start time of chain $\job{1}$ (resource's offset is $o_2=40$), and we stack each interval of length $w=40$ on top of the previous one. So row $v$ describes the schedule of interval $[v\cdot w, (v+1)w)$. 
Horizontal coordinate $u$ corresponds to the relative position of the task's occurrence in the interval.

 Row indices are represented by two-digit numbers using a mixed-radix system based on the period multipliers 2 and 3: here period $\per{2}=2\cdot\per{1}$, and $\per{3}=3\cdot\per{2}$.
 
Figure~\ref{fig:sample_placement} shows the corresponding height-divisible packing view of the same solution. Rectangles are described with coordinates of their bottom-left corner $(x,y)$. 
The packing is obtained by interchanging the rows as follows:
we swap the two-digit row numbers in  Figure~\ref{fig:sample_hyperperiod} such that the most significant digit becomes the least significant digit and vice versa. Finally, we obtain Figure~\ref{fig:sample_placement} by ordering the rows according to their new two-digit numbers.
For example, the row with $\job{6}$ has number $0_3 1_2 = 1$ in Figure~\ref{fig:sample_hyperperiod} and after the swap, the same row has number $1_2 0_3 = 3$ in Figure~\ref{fig:sample_placement}. We can see the occurrences of the original tasks have been merged, and the dimensions of the resulting rectangles correspond to the rectangles of the associated packing instance. For example, tasks 1-2 with the shortest period are transformed into rectangles spanning the entire height of the bin. Tasks 3-7 merge three of their occurrences into a rectangle with height 3, and finally, rectangles associated with the longest-period tasks 8-10 have a height equal to 1.

Note that thanks to the equivalence between the two problems, we can recover a periodic schedule by providing a solution for the height-divisible packing problem and vice versa.

\section{Complexity Analysis}
\label{sec:complexity-analysis}
In this section, we consider related complexity results to explain the complexity of our problems. 
First, we know that scheduling periodic tasks with harmonic periods on a single resource is a strongly NP-hard problem \cite{cai1996nonpreemptive} whereas it becomes polynomial if processing times and periods belong to the harmonic set \cite{korst1996scheduling}. So, minimizing one of our objectives seems at least as difficult as this problem and is still NP-hard unless very strong assumptions are made. In the rest of the section, we show some special polynomial cases of our problem and present a strongly NP-hard subproblem.
\subsection{Two Resources with Chains of Length Two.}
In this subsection, we assume that there are only two resources and that chains either start with a task on resource 1 and end with a task on resource 2, or vice versa.
\subsubsection{Upper Bounds for the Two Resource Case}
We prove that if we build a schedule as in \cref{algobase} from a core schedule on each resource, then the degeneracy is bounded.
\begin{property}Assume that $\job{k}$ is a chain of length two on two different resources. A schedule $\first{}{}$ built according to \cref{algobase} satisfies: if $\proctime{1}{k}+\proctime{2}{k}\le \perjob{k}$ then $\DGk{k}\le 1$ else $\DGk{k}\le 2$.
\label{prop:2machines}
\end{property}
\begin{proof}
    We consider a chain $\job{k}$, and we assume without loss of generality that it starts on the first resource. 
    Let us consider $\sigma$ a core schedule satisfying the resource constraints, and assume that $\sigma_i^k<\perjob{k}$ for $i\in\{1,2\}$. We can observe that due to the fact that the processing time of each task is not greater than $\perjob{k}$, we have the following inequalities:
    \begin{equation}
    0\le \proctime{1}{k}+\proctime{2}{k}<2\perjob{k}
        \label{eq:inegalities}
    \end{equation}

    Notice that in any case of the construction given in \cref{algobase}, $\first{1}{k}=\sigma_1^k$. Let us consider the different cases of computation of $\first{2}{k}$ applied in \cref{algobase}. For each case, we show the bounds of degeneracy.

    Case 1: $\sigma_1^k+\proctime{1}{k}\le \sigma_2^k$. In this case \cref{algobase} sets $\first{2}{k}=\sigma_2^k$. 
    \begin{eqnarray}\SE{k}= \sigma_2^k+\proctime{2}{k}-\sigma_1^k<\sigma_2^k+\proctime{2}{k}\\
    \Rightarrow \SE{k}<2\perjob{k}\quad
    \Rightarrow\quad  \DGk{k}\le 1\end{eqnarray}

    Case 2: $\sigma_2^k< \sigma_1^k+\proctime{1}{k}\le \sigma_2^k+\perjob{k}$. Then \cref{algobase} sets $\first{2}{k}=\sigma_2^k+\perjob{k}$.
 By~\cref{eq:inegalities}:
\begin{eqnarray}
\SE{k}=\sigma_2^k+\proctime{2}{k}+\perjob{k}-\sigma_1^k<\proctime{1}{k}+\proctime{2}{k}\\
\Rightarrow\SE{k}<2\perjob{k}\quad 
\Rightarrow \quad \DGk{k}\le 1
\end{eqnarray}
Notice that in the special case where $\proctime{1}{k}+\proctime{2}{k}\le\perjob{k}$, then $\DGk{k}=0$.

Case 3:  $\sigma_2^k+\perjob{k}< \sigma_1^k+\proctime{1}{k}$. Then \cref{algobase} sets $\first{2}{k}=\sigma_2^k+2\perjob{k}$. In this case:
\begin{eqnarray}
\sigma_2^k+\proctime{2}{k}-\sigma_1^k<\proctime{1}{k}+\proctime{2}{k}-\perjob{k}\\
\Rightarrow\SE{k}=\sigma_2^k+\proctime{2}{k}+2\perjob{k}-\sigma_1^k<\proctime{1}{k}+\proctime{2}{k}+\perjob{k}
\end{eqnarray}
Now in any case from \cref{eq:inegalities}, 
\begin{equation}
    \SE{k}< 3\perjob{k}\quad\Rightarrow \quad \DGk{k}\le 2
\end{equation}

Notice that in the special case where $\proctime{1}{k}+\proctime{2}{k}\le \perjob{k}$, then we would have $\SE{k}\le 2\perjob{k}$, so that $\DGk{k}\le 1$. 
    
\end{proof}
\begin{corollary}
 In the case of processing times and periods belonging to a harmonic set, building a schedule on two resources of chains of length two with $\DGm\le 1$ can be done in polynomial time.
\end{corollary}
\begin{proof}
    Notice that if processing times are harmonic, then the processing time of a task is at most half of the period. Hence, we can easily establish that in this case, $\proctime{1}{k}+\proctime{2}{k}\le\perjob{k}$. This implies that building two independent periodic schedules on each resource in polynomial time with the algorithm of \cite{korst1996scheduling} and then applying \cref{algobase} leads to a schedule in which for each chain $\job{k}$, $\DGk{k}\le 1$ according to Property~\ref{prop:2machines}.
\end{proof}
 \subsubsection{Scattering on a Line of Two Resources with a Single Period}\label{sec:single-per}
Let us consider a special case of the line support graph with two resources and a scattering hypothesis. So, all chains have their first task on resource 1 and second on resource 2. Assume all chains have the same period $T$. Notice that in this case, the sum of processing times of tasks on any resource is not greater than $T$. Otherwise, the problem would be infeasible. Then, finding a feasible schedule for a single resource becomes trivial. We can prove that as for the two-machine flow-shop problem, minimizing the degeneracy can be done in polynomial time. The same is true for deciding the existence of a periodic schedule with no wait time.

\begin{property}
For any instance of a scheduling problem $PD2|T_i=T,jit=0, chains|\DGs$, in the special case of scattering  on a line of two resources:
\begin{itemize}
    \item Assuming that the sum of processing times on each machine is not greater than period $T$, there always exists a feasible schedule with \begin{equation}
        \DGs=\DGm=0
    \end{equation}
    \item It can be decided in polynomial time whether exists a feasible schedule with a no-wait hypothesis (i.e., where $\SE{k}=\proctime{1}{k} + \proctime{2}{k}$  for each chain $\job{k}$)
\end{itemize}
\label{lem:2proc}
\end{property}
\begin{proof}
The non-periodic version of this problem is a two-machine flow-shop. A chain $\job{k}$ corresponds to a job with two operations $\task{1}{k},\task{2}{k}$ performed on the first and second machine, respectively. As shown in \cite{Johnson54}, a schedule that optimizes the makespan ($C_{max}=\max_{k\in\{1,\ldots,\kappa\}}\first{2}{k} +\proctime{2}{k}$) can be computed in polynomial time using Johnson's rule. Moreover, we know that it is a permutation schedule: the order of jobs is the same on both machines even though it was not required. 

If we shift the schedule of the first machine to the left and of the second machine to the right so that there does not remain idle time between two tasks on the resources and it ends at the optimal makespan, then the schedule can be repeated every $T$ time units without conflicts. Let $\first{}{}$ be the Johnson's schedule shifted as indicated above.

We prove that this leads to a zero degeneracy periodic schedule.
Indeed, consider a pivot chain $\job{k^*}$, for which $\first{1}{k^*}+\proctime{1}{k^*}=\first{2}{k^*}$. Such a pivot chain exists. Otherwise, the tasks on the second machine could be scheduled a little earlier, and thus, the makespan would not be optimal. The start-to-end latency for chain $\job{k^*}$ is thus minimal, and so is its degeneracy. Then, for any chain $k$ performed before this $k^*$ in the schedule, the start-to-end latency $\SE{k}\le \first{2}{k^*}$   is not greater than the sum of processing times on the first resource, which is not greater than $T$. Hence $\DGk{k}=0$. Symmetrically, for a chain $l$ performed after chain $k^*$, the value of $\SE{l}$ should be less than the sum of processing times on the second resource, which is not greater than $T$ so that $\DGk{l}=0$. This proves that in this schedule, we have $\DGm= \DGs=0$.

Now consider the no-wait problem where we wish that for each chain $\job{k}$ the start-to-end latency is $\SE{k}=\proctime{1}{k}+\proctime{2}{k}$. As proved in \cite{Reddi01091972}, a  no-wait periodic schedule with minimal period $T'$ that minimizes the idle time on the second machine  can be built using a polynomial subcase of the traveling salesman problem described 
in \cite{gilmore1964sequencing}. So if the period $T$ is not less than $T'$, we can just delay the iterations to get a periodic schedule with period $T$. Otherwise, the problem is not feasible.
\end{proof}

Property~\ref{lem:2proc} indicates that minimizing $\DGs$ and $\DGm$ is polynomial on two machines with a single period. Still, the complexity of minimizing  $\sum_k\SE{k}$ or $\max_k\SE{k}$ are open problems and might be more difficult if we consider that flow-shop problem on two machines with limited buffers is NP-hard problem (see survey \cite{HallS96}).

\subsubsection{Scattering on a Line of Two Resources with Two Periods}

As in \cref{sec:single-per}, we assume scattering chains on two resources, but now the chains can have two different harmonic periods. In \cite{korst1996scheduling}, the authors give a reduction from  $3$-PARTITION to a single resource scheduling problem with two different harmonic periods, thus proving that the feasibility of single resource PSP with two periods is strongly NP-hard. 

From this result, we can deduce the following:

\begin{property}
    For scattering chains on two resources with two harmonic periods on a line, finding a schedule with $\DGs$ or $\DGm$ equal to zero is strongly NP-hard.
    \label{prop:2proc2per}
\end{property}
\begin{proof}
Consider an instance of the single resource scheduling problem with two harmonic periods. We build a simple reduction to the problem on two resources. We are given a set $T$ of tasks, a task $\task{}{k}
\in T$ has period $\perjob{k}\in\{\per{1},\per{2}\}$, a set of two harmonic periods. Notice that the processing time of $\task{}{k}$ can be assumed strictly less than its period; otherwise, $\task{}{k}$ would be the only task scheduled on the resource. For each task $\task{}{k}
\in T$, we define a chain $\job{k}$ with period $\perjob{k}$ composed of task $\task{1}{k}=\task{}{k}$ on the first resource and a task $\task{2}{k}$ with processing time $1$ on the second resource. Obviously, if a feasible schedule for the single resource instance exists, it gives a schedule for the two resources of the chains with $0$ degeneracy by scheduling the tasks on the second resource as soon as possible with respect to precedence constraint. Conversely, a schedule with $0$ degeneracy $\DGs$ or $\DGm$ implies the existence of a feasible schedule on the first resource.
So,  scattering problem on two resources with two periods is NP-hard in the strong sense.
\end{proof}

As was shown in \cite{korst1996scheduling}, the single resource problem with processing times and the periods belonging to a harmonic set is polynomial. However, the complexity of building zero-degeneracy schedules for the two-resource scattering on a line where periods and processing times are from one harmonic set remains open.

\subsection{Acyclic Instance with Bottleneck Resource and Equal Processing Times}
\label{sec:line-tree-top}
In this section, we assume equal processing times of all tasks of a chain and that the support graph is acyclic and contains a bottleneck resource $b$. We first consider the special case of a line support graph and then extend it to other graph structures.

\subsubsection{Bottleneck Resource along a Line}\label{sec:offset-schedule}
We consider a case when messages are sent in a single direction with a line support graph and a single bottleneck resource and equal processing times of all tasks of a chain. More formally, we consider a support graph that forms a unique directed path from resource $1$ to resource $M$ and a bottleneck resource $b\in\{1,\ldots, M\}$. We consider that all chains $\job{k}$ have a source resource $u_k\le b$ that performs the first task of the chain ($\pi_{1}^k=u_k$) and a sink resource $v_k\ge b$ that performs the last task of the chain ($\pi^k_{n^k}=v_k$), and if $\task{i}{k}$ is performed on some resource $m$, with $i<n^k$, then $\task{i+1}{k}$ is performed on resource $m+1$.  
Notice that scattering corresponds to the case where $b=1$. Let us denote by $\proctime{}{k}$ the processing time of every task of chain $\job{k}$. 

Let us assume that we have a schedule $\sigma$ for resource $b$, which contains one task of each chain. We define a special kind of schedules, called offset schedules, that define schedules of other resources according to $\sigma$, with an offset depending only on the resource.

\begin{definition}[Offset schedule] An offset schedule associated with  a schedule $\sigma$ on resource $b$ is a schedule built by assuming that for any chain $\job{k}$ and any resource $m\in\{u_k,\ldots,v_k\}, m\neq b$,  a $\task{i}{k}\in\tasksmachine{m}$ is scheduled exactly as it is in resource $b$, with an offset $o_m$ that only depends on $m$, and can be negative or positive. If $\task{j}{k}$ is the task of chain $\job{k}$ performed on resource $b$, the start time of $\task{i}{k}$ is thus  $\first{i}{k}=\sigma_j^k+o_m$. 
\end{definition}

Notice that as long as $\sigma$ satisfies the constraint on resource $b$, any offset schedule based on $\sigma$ also satisfies the resource constraints on the other resources. The start-to-end delay of a chain $\job{k}$ in an offset schedule is $\SE{k}=\proctime{}{k}+o_{v_k}-o_{u_k}$.
 The next property establishes that an offset schedule with an approximation ratio can be built.
 
\begin{property}
Suppose that all chains start before a bottleneck resource $b$ and end after $b$, along a line, assuming equal processing times within each chain.  
Suppose we are given a feasible schedule $\sigma$ of the tasks on the bottleneck resource $b$.
The minimal degeneracy $\DGm$ or $\DGs$ of an offset schedule $\omega$ associated with $\sigma$ can be computed in polynomial time. In this schedule, we have the following inequality:
\begin{equation}
    \DGm(\omega)\le \frac{p^{\max}}{p^{\min}}(\DGm(opt)+1),
\end{equation}
where $p^{\min}=\min_{k}\proctime{}{k}, p^{\max}=\max_{k}\proctime{}{k}$ and $\DGm(opt)$ is the optimal $\DGm$.
    \label{prop:bottleneck}
\end{property}

\begin{proof}
We first establish the necessary conditions for offsets to give a feasible schedule; then, we prove that we can determine the offsets so that the start-to-end delay of chains is minimal (among offset schedules). Eventually, we show the performance ratio by using a lower bound on the star-to-end latency.

  Precedence relations imply that in any feasible offset schedule based on schedule $\sigma$, we have the following inequality: \begin{equation}\text{If} \ m>b, \quad o_m-o_{m-1}\ge \max_{k,v_k\ge m}\proctime{}{k}\end{equation} Indeed, if $\task{i}{k}$ is the task of chain $\job{k}$ performed on resource $m$ ($v_k\ge m$), then due to the line support graph $\task{i-1}{k}$ is performed on resource $m-1$. Hence $\first{i}{k}\ge \first{i-1}{k}+\proctime{}{k}$, which proves that: 
  \begin{equation}
      o_m-o_{m-1}\ge \proctime{}{k}
  \end{equation}
   With similar arguments, we can prove that: \begin{equation}\text{If}\ m<b,\quad o_{m+1}-o_{m}\ge \max_{k,u_k\le m}\proctime{}{k}\end{equation} Indeed, if $\task{i}{k} $ is performed on resource $m$ ($u_k\le m$), $\task{i+1}{k}$ is performed on resource $m+1$, so that $\first{i+1}{k}\ge \first{i}{k}+\proctime{}{k}$. So: 
   \begin{equation}
      o_{m+1}-o_{m}\ge \proctime{}{k}
  \end{equation}.
  
  We can now build a feasible offset schedule $\omega$ that minimizes the start-to-end latency of each chain (among offsets schedules) by setting $o_b\coloneqq0$ and if $m>b$ then $o_m\coloneqq o_{m-1}+ \max_{k,v^k\ge m}\proctime{}{k}$, whereas if $m<b$, then $o_{m}\coloneqq o_{m+1}-\max_{k,u^k\le m}\proctime{}{k}$. This can be computed in time complexity $\mathcal{O}( \mu log(\mu))$ by sorting the processing times and, for at most $M$ steps (from $b$ to $M$ and from $b-1$ to $1$), scanning the list to find the maximum processing time concerned by resource $m$.
   
  Then, for a chain $\job{k}$ it holds: 
  \begin{equation}
      o_{v_k}-o_b\le (v_k-b)p^{max},
  \end{equation} whereas:
  \begin{equation}
      o_{b}-o_{u_k}\le (b-u_{k})p^{max}.
  \end{equation}
  
  The start-to-end latency of $\job{k}$ in this schedule satisfies: 
  \begin{equation}
      \SE{k}(\omega)\le o_{v_k}+\proctime{}{k}-o_{u_k}\le (v_k-u_k+1)p^{max}
  \end{equation}
  
To conclude, we observe that the start-to-end latency of chain $\job{k}$ in any schedule is not less than $(v_k-u_k+1)p^{min}$. 
 So, for any schedule $s$, $\displaystyle \frac{\SE{k}(\omega)}{\SE{k}(s)}\le\frac{p^{\max}}{p^{\min}}$.

Now, for some chain $\job{k}$:
\begin{equation}
    \DGm(\omega)=\left\lceil\frac{\SE{k}(\omega)}{\perjob{k}}\right\rceil -1\le \frac{\SE{k}(\omega)}{\perjob{k}}\le\frac{p^{\max}}{p^{\min}}\frac{\SE{k}(opt)}{\perjob{k}}\le 
\frac{p^{\max}}{p^{\min}}(\DGm(opt)+1)
\end{equation}
\end{proof}

In the case of processing times and periods belonging to a harmonic set, a schedule $\sigma$ of the bottleneck resource can be found in polynomial time. Then Property~\ref{prop:bottleneck} gives a polynomial approximation algorithm in this case, assuming an acyclic line support graph with a bottleneck resource, equal processing times within chains, and harmonic processing times and periods.

When processing times are not within a harmonic set with the periods, we can still build an offset schedule as long as we are able to find a feasible schedule using, e.g., CP. First,
we find a feasible schedule for the first resource of the line, and then we create
schedules for other resources by applying a suitable offset. Even though
the single resource problem remains strongly NP-hard, we empirically show in
\cref{sec:exp-line} that the offset schedule combined with the CP model for the single
resource problem works very well.

\subsubsection{Bottleneck Resource on Any Acyclic Graph}
We now consider that the support graph $G=(V,E)$ does not contain any directed cycle and that there is a bottleneck resource $b$. We show that the previous result can be extended to this case. As previously, we assume a feasible schedule $\sigma$ on bottleneck resource $b$. Two consecutive tasks of chain $\job{k}$, $\task{i}{k}, \task{i+1}{k}$ with $\pi_i^k=m$, $\pi_{i+1}^k=m'$ then satisfy that $(m,m')\in E$ is an arc of the support graph. We say that the chain passes through arc $(m,m')$. For an arc $a=(m,m')\in E$ of the support graph, let us define by $C(a)$ the set of chains $\job{k}$ that pass through arc $a$.

\begin{figure}
    \centering
    \includegraphics[width=0.9\linewidth]{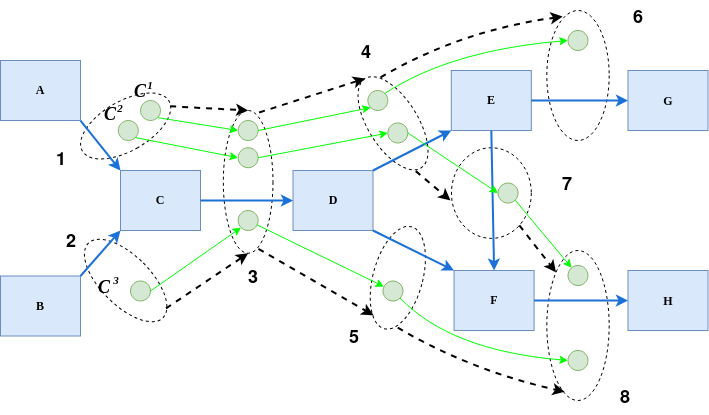}
    \caption{Example of acyclic PSP with a bottleneck resource. Underlying network topology (blue) induces an appropriate support graph (dashed) given the three chains of tasks (green) starting at links A-C and B-C, passing through bottleneck link C-D.}
    \label{fig:linkstree}
\end{figure}

In \cref{fig:linkstree}, we present an example of an instance with an acyclic support graph and a bottleneck resource 3. Notice that the support graph contains an undirected cycle (resources 3, 4, 5, 7, 8). This does not invalidate the results of this section since there is still no directed cycle. 

A crucial distinction between the network topology and the support graph can be observed in \cref{fig:linkstree} by focusing only on network nodes A, B, and C (and removing nodes D-H). In that case, even though the messages share the sink network node C, the support graph, consisting only of resources 1 and 2, does not contain any edges (and no bottleneck resource). Thus, such PSP decomposes into scheduling chains $\job{1},\job{2}$ on resource~1 and independently scheduling chain $\job{3}$ on resource~2.

In the remainder of the subsection, we do not make other specific assumptions on the support graph's structure. However, we will show that the performance of offset schedules can be measured with respect to a parameter $\eta$.
\begin{definition}[Ratio between path lengths] Parameter $\eta$  measures the maximum ratio between the length of two different paths between two nodes in the support graph $G$. More formally, if $P$ is a directed path of the support graph from node $m$ to node $m'$, then its length $\lambda(P)$ is the number of nodes of the path, and $\mathcal{P}(m,m')$ denotes the set of paths from $m$ to $m'$ in $G$. We also denote by $m\rightarrow m'$ the fact that $\mathcal{P}(m,m')\neq\emptyset$. Then
\begin{equation}
    \eta=\max_{m,m'\in V, m\rightarrow m'}\frac{\max_{P\in\mathcal{P}(m,m')}\lambda(P)}{\min_{P\in\mathcal{P}(m,m')}\lambda(P)}
\end{equation}
\end{definition}

Notice that in several important special cases, for example, line and tree support graphs (or when any two paths from $m$ to $m'$ have the same length), we have $\eta=1$. This does not hold for \cref{fig:linkstree}, where there are multiple paths between resources 3 and 8. 
As previously mentioned, an offset schedule will define the schedule of resource $m$ based upon $\sigma$ with an offset $o_m$, assuming offset of the bottleneck resource $o_b=0$. We can now prove the following property of offsets:
\begin{lemma}
For any arc $a=(m,m')\in E$ of the support graph, in any offset schedule:
\begin{equation}
    o_{m'}-o_m\ge \max_{k, \job{k}\in C(a)}\proctime{}{k}
\end{equation}
    \label{lem:offestgraph}
\end{lemma}
\begin{proof}
Suppose that a chain $\job{k}\in C(a)$. Then $\task{i}{k}, \task{i+1}{k}$ are consecutive in the chain with $\pi_i^k=m,\pi_{i+1}^k=m'$. Suppose that $\task{j}{k}$ is the task of chain $\job{k}$ performed on resource $b$. The precedence constraint between $\task{i}{k}, \task{i+1}{k}$ implies that: \begin{equation}o_m+\sigma_j^k+\proctime{}{k}\le o_{m'}+\sigma_j^k.\end{equation}
\end{proof}

So we can now consider the support graph $G$ with its arc valuation $v$: for each arc $a=(m,m')\in E$ $v(a)=\max_{k, \job{k}\in C(a)}\proctime{}{k}$. The valuation $v(P)$ of a path $P$ of $G$ is defined as the sum of the valuation of its arcs. From this valuation we define the following offsets $\hat{o}_m$:
\begin{definition}[Maximal path offsets $\hat{o}$]
If there is a path from $b$ to $m$ in G, then we define :
\begin{equation}
    \hat{o}_m=\max\{v(P)|P\ path\ from\ b\ to\ m \}
\end{equation}
Otherwise:
\begin{equation}
    \hat{o}_m=\max\{-v(P)|P\ path\ from\ m\ to\ b \}
\end{equation}
\end{definition}
\begin{thm}
    In the case of acyclic instance with a bottleneck resource $b$, and given a schedule $\sigma$ of resource $b$, the offset schedule $\omega$ defined upon maximal path offsets $\hat{o}_m$ can be computed in polynomial time from $\sigma$ and has the property that
    \begin{equation}
        \DGm(\omega)\le\eta\cdot \frac{p^{\max}}{p^{\min}}(\DGm(opt)+1),
    \end{equation}
    where $p^{\min}=\min_{k}\proctime{}{k}, p^{\max}=\max_{k}\proctime{}{k}$ and $\DGm(opt)$ is the optimal $\DGm$.
    \label{thm:offsetgraph}
\end{thm}
\begin{proof}
We first prove that any offset schedule satisfies that the offset difference along a directed path induced by a chain $\job{k}$ gives a potential inequality. Then, we prove that offsets $\hat{o}$ satisfy these inequalities and can be computed in polynomial time. We then measure the start-to-end latency of a chain in schedule $\omega$ and give a lower bound on the start-to-end latency of a chain that induces our approximation ratio for the degeneracy.

  Let us consider a chain $\job{k}$ starting on resource $m=\pi_1^k$ and ending at resource $m'=\pi^k_{n^k}$. Since $b$ is a bottleneck resource, $\job{k}$ induces a directed path $P$ from resource $m$ to $b$ in $G$ and a path $P'$ from $b$ to $m'$. By   Lemma~\ref{lem:offestgraph}, summing the inequalities on the paths implies that $o_{m'}-o_b\ge v(P')$ whereas $o_b-o_m\ge v(P)$. By definition, these inequalities are satisfied by $\hat{o}$ values. Paths with maximum value can be computed in polynomial time ($\mathcal{O}(M^2)$ in the acyclic graph $G$). Now let us consider the start-to-end latency of chain $\job{k}$ for schedule $\omega$. 
   As in Property~\ref{prop:bottleneck} we get $\SE{k}(\omega)=\hat{o}_{m'}+\proctime{}{k}-\hat{o}_{m}$.
   Now observe that the maximum value of a path $P$ from $m$ to $m'$ plus the processing time $\proctime{}{k}$ is not greater than $\max_{Q\in\mathcal{P}(m,m')}\lambda(Q)p^{\max}$, whereas the optimal start-to-end latency of chain $\job{k}$ is not less than $(n^k+1)p^{min}\ge \min_{Q\in\mathcal{P}(m,m')}\lambda(Q)p^{\min}$. So we get \begin{equation}
       \SE{k}(\omega)\le \eta\cdot\frac{p^{\max}}{p^{\min}}\SE{k}(opt)
   \end{equation}
   We can then use the same arguments as in the proof of Property~\ref{prop:bottleneck}  to conclude.
\end{proof}

\begin{corollary}
    In the case of a tree support graph and a bottleneck resource $b$, and processing times and periods belonging to a harmonic set, a schedule $\omega$ with the following approximation ratio can be computed in polynomial time. $$\DGm(\omega)\le\frac{p^{\max}}{p^{\min}}(\DGm(opt)+1)$$
\end{corollary}
\begin{proof}
    We just have to build, using Korst's algorithm \cite{korst1996scheduling}, a schedule $\sigma$ for the bottleneck resource and then compute the optimal offset schedule associated with $\sigma$ along with Theorem~\ref{thm:offsetgraph}.
\end{proof}

\section{Solution Approach and Algorithms}
\label{sec:methods}

\begin{figure}[htbp]
\centering
\includegraphics[width=0.99\textwidth]{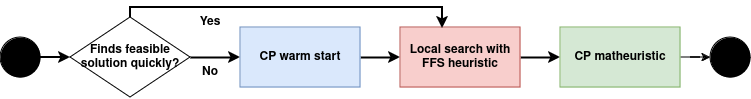}
\caption{Outline of the proposed approach.}
\label{fig:method-diagram}
\end{figure}

This section describes the proposed solution approach for solving PSPs with harmonic periods. The approach is based on local search \cite{hladik2020complexity}, which uses the first fit scheduling (FFS) heuristic. The overall flow, visualized in \cref{fig:method-diagram}, shows the local search and two CP-based improvements. If the local search (\cref{sec:heuristic}) is unable to find a feasible solution quickly, the first improvement provides it with the warm start (\cref{sec:warmstart}); this is particularly useful for solving highly-utilized instances. The second improvement reduces the degeneracy of the found schedule even further using a matheuristic (\cref{sec:matheuristic}).

\subsection{Local Search}\label{sec:heuristic}
The local search method is outlined in \cite{hladik2020complexity}. The inner FFS heuristic processes the tasks following the ordered list of tasks $\mathcal{\orderTasks}$; the ordered lists are generated and supplied to the heuristic by the encompassing local search. One by one, the tasks are scheduled on their dedicated resources as shown in \cref{alg:first-fit}.

We distinguish two methods. The \emph{leftmost} method schedules each task as early as possible so that none of its occurrences overlap with any occurrence of any already scheduled tasks (for more details, see Section 4 in \cite{hladik2020complexity}). This method is more suitable for finding a feasible solution, but it may result in a large degeneracy since it ignores the precedence constraint. The second \emph{predecessor} method performs the following: If predecessor $i$ of the task $i+1$ in chain $k$ is already scheduled, we schedule the current task only after the completion time of its predecessor. Unlike the previous method, the \emph{predecessor} method does not ignore the precedence of already scheduled predecessors. Note that by scheduling the successor task after alternatively selected time $t_i^k + l_{i,i+1}^k$, we could modify the algorithm to consider both link delays ($l_{i,i+1}^k > \proctime{i}{k}$) as well as cut-through mechanisms ($l^k_{i,i+1} < \proctime{i}{k}$). In both cases, the actual schedule is obtained by postponing the tasks that do not adhere to the precedence constraints, according to \cref{postpone-lemma}.

\begin{algorithm}
\DontPrintSemicolon
\SetAlgoLined
\SetNoFillComment
\LinesNotNumbered 
 \SetKwInOut{Input}{Input}
\SetKwInOut{Output}{Output}

\Input{Tasks $\task{i}{k}$ with periods, processing times, and assignment to resources; chains $\job{k}$ with precedences; method; ordered list $\mathcal{\orderTasks}$}
\Output{Schedule; or infeasible $\mathcal{\orderTasks}$}
\;
 \While{$\mathcal{\orderTasks} \neq \{\}$}{
  $\task{i}{k} \gets \mathcal{\orderTasks}$.pop() \tcp{remove first task from the ordered list}
  \eIf{$\emph{method}$ $=$ {predecessor}}{
       $\task{i-1}{k} \gets $ predecessor of $\task{i}{k}$
       
       \eIf{$\task{i-1}{k}$ exists and already scheduled}{ 
   $t\gets \first{i-1}{k} + \proctime{i-1}{k}$}{$t \gets 0$}
 }{$t \gets 0$ }

schedule $\task{i}{k}$ as early as possible starting from $t$

 \If{$\task{i}{k}$ is not scheduled}{\Return infeasible $\mathcal{\orderTasks}$}
 }

\ForEach{$\job{k}$} 
{
\tcp{postpone tasks to satisfy precedences}
\For{$i\leftarrow 1$ \KwTo $n^k - 1$}{
    $q \gets \min q~|~t_{i+1}^k + q\cdot \perjob{k} \ge t_{i}^k + \proctime{i}{k}$
    
    $t_{i+1}^k \gets t_{i+1}^k + q\cdot \perjob{k}$
}
} 
 \Return schedule
 \caption{First fit scheduling heuristic}
 \label{alg:first-fit}
\end{algorithm}

We use local search to find the appropriate ordered list $\mathcal{\orderTasks} = (\task{1}{}, \dots, \task{N}{})$, where $N$ is a total number of tasks, which is provided to the FFS heuristic. Given an initial ordered list, the local search first tries to modify it to respect the precedences of chains, one chain $k$ at a time, e.g.: 

\begin{equation}
\label{eq:neigh}
    (\dots, \task{2}{k}, \task{3}{k}, \dots, \task{1}{k}, \dots) \rightarrow (\dots, \task{1}{k}, \task{2}{k}, \dots, \task{3}{k}, \dots)
\end{equation}

Once this fails, the local search performs either random moves or chain reordering, each with 50 \% probability. The random moves are performed by swapping two tasks in the ordered list.

\begin{equation}
    (\dots, \task{i}{k}, \dots, \task{j}{l}, \dots) \rightarrow (\dots, \task{j}{l}, \dots, \task{i}{k}, \dots)
\end{equation}

When a random move is performed, the tasks to be swapped are either two arbitrary tasks, two tasks of the same chain ($k=l)$, or two tasks that are consecutive in the given chain ($k=l \wedge j = i+1)$. The variant of the random move is selected uniformly random. The reordering operation takes one (arbitrary) chain whose order of tasks in the list does not follow its precedence constraints and reorders these tasks to follow them, as shown in \cref{eq:neigh}. 

In the first iteration, the local search begins with a rate-monotonic ordered list, which sorts tasks in ascending order of their periods. As discussed in \cite{grus2024icores}, the rate-monotonic ordered list is the desirable starting point since it processes the most constrained tasks (i.e., tasks with the most occurrences) at the start of the procedure when the schedule is mostly empty.

Finally, this local search procedure could easily be used with simulated annealing, tabu search, or another metaheuristic scheme. However, based on the results of \cite{hladik2020complexity} and limited experimentation, we utilize the pure local search and try to alleviate its possible disadvantages using the CP approaches described in the following sections.

\subsection{Warm-Start by Constraint Programming }\label{sec:warmstart}
The local search generally works very well. However, with the utilization of instances close to 100\%, the approach struggles to find any feasible solution. In that case, it is beneficial to provide the local search with an alternative initial ordered list, which the heuristic maps to a feasible schedule immediately.

\subsubsection{Feasible Schedule for a Single Resource}
We use the bin-packing CP formulation for monoprocessor (as described in \cite{grus2024icores}) based on the height-divisible packing of \cref{sec:packing-section}. The CP solver processes one resource at a time and creates a feasible schedule for each. Afterward, these feasible schedules are used to infer a suitable ordered list for the local search.

\begin{figure}
    \centering
    \includegraphics[width=0.8\linewidth]{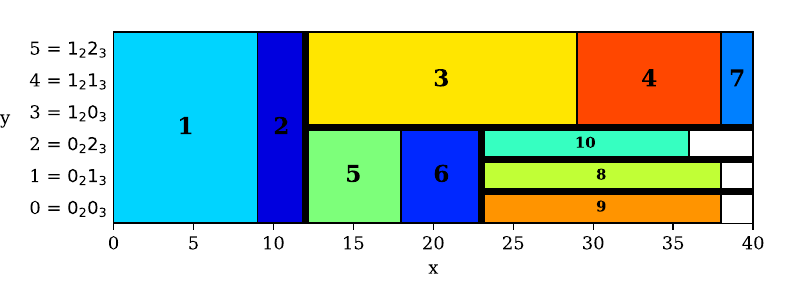}
    \caption{Canonical form HD2D packing obtained by CP warm start. Bold T-like lines show how the bin is recursively partitioned into sub-bins.}
    \label{fig:sample_canonical}
\end{figure}

A set of tasks $\tasksmachine{m}$ (with periods $\per{1},\dots,\per{j},\dots,\per{\rho}$) induces an associated set of rectangles $\mathcal{R}_m$, whose heights and widths are defined in  \cref{sec:packing-section}. The set of available heights corresponds to the set of periods of the tasks: $h_j=\per{\rho}/\per{j}$. Thus, $h_1 = H$, which is equal to the height of the entire bin, and $h_\rho=1$. In \cref{fig:sample_canonical}, we see that there are two rectangles that span the entire height of the bin and three rectangles of height 1. These are induced by single-resource PSP outlined in \cref{fig:sample_hyperperiod}.

The formulation packs the rectangles of $\mathcal{R}_m$ in a height-divisible manner to produce a packing in canonical form. This means that for each rectangle, it holds that all the rectangles packed to its left have a height greater or equal. \cref{fig:sample_canonical} shows such packing (unlike in \cref{fig:sample_placement}). It was proven in \cite{grus2024icores}) that if there is a feasible packing, then there exists a feasible canonical packing. 

The T-like bold lines in the figure show how the bin is recursively partitioned into sub-bins, where rectangles with smaller heights can be packed, given the height-divisible property of packing. Rectangles with height $h_1=H=6$ can only be packed into a single sub-bin that spans the entire figure. The rectangles with height $h_2 = 3$ have two available sub-bins. Finally, the rectangles with $h_3=1$ can be packed into six sub-bins. However, since only three of these sub-bin of height 1 are filled with rectangles, the other three have width 0 and are not shown in \cref{fig:sample_canonical}. 

To utilize this sub-bin view of the packing, we partition the rectangles to lists $\mathcal{R}_m^{j}$ according to their height $h_j$. Each such list is associated with a vector of widths of the rectangles $\mathbf{l}^j$ and a vector of integer assignment variables $\mathbf{a}^{{j}}$ with domain $\{0,\cdots,(H/h_j)-1\}$. The assignment variable determines which sub-bin the rectangle will be packed into. In \cref{fig:sample_canonical}, height $h_2=3$ is relevant for rectangles $\mathcal{R}_m^{2} = (3, 4, 5, 6, 7)$, with widths $\mathbf{l}^2 = (17,9,6,5,2)$. The top sub-bin for $h_2=3$ is occupied by rectangles  3, 4, and 7. The bottom sub-bin for $h_2=3$ is occupied by rectangles  5 and 6. Therefore, the vector of assignment variables is $\mathbf{a}^{{2}}=(1,1,0,0,1)$. 

Furthermore, we create a vector of integer variables $\mathbf{L}^{j} = (L^{j}_0,\dots,L^{j}_k,\dots, L^{j}_{(H/h_j)-1})$ for each height $h_j$. This vector contains a single variable for each available sub-bin $k$ of height $h_j$. This load variable keeps track of how much the sub-bin is filled. For example, the load variable $L^1_0$, associated with the only sub-bin of height $h_1=6$, is set to 12 in \cref{fig:sample_canonical}. The sub-bins for $h_2=3$ are filled $L^2_0=11$, $L^2_1=28$ respectively. The width of the bin $w$, into which we pack the rectangles, is given by the shortest period $\per{1}$.

 The height-divisible bin-packing problem is modeled as follows:
\begin{align}
	& \mathtt{Pack}(\mathbf{L}^{j}, \mathbf{a}^{j}, \mathbf{l}^{j}) & \forall j \in \{1,\dots,\rho\} \label{eq:pack}\\
    &\sum_{j=1}^{\rho} L^{j}_{\lfloor k / h_{j}\rfloor} \le w & \forall k \in \{0,\dots, H-1 \} \label{eq:pack2}  
\end{align}

The well-known global pack constraint \cite{Shaw2004cpconstraint} (\cref{eq:pack}) forces the model to assign to each individual variable of $\mathbf{a}^{{j}}$ an index $k$ of one of the sub-bins with load $L^{j}_k$. It also sets the value of $L^{j}_k$ to the sum of the widths of assigned rectangles, thus solving a bin-packing subproblem for each distinct height $h_j$. 

The second constraint \cref{eq:pack2} enforces that for each row of height 1, the sum of loads of sub-bins that cross the row (across the different heights) does not exceed the bin's width. In case of row 2 in \cref{fig:sample_canonical}, we have:
\begin{equation}
    L^1_0 + L^2_0 + L^3_2 = 12 + 11 + 13 = 36 \le 40 = w
\end{equation}

Once the CP solver finds a feasible HD2D packing in canonical form (given the values of assignment variables $\mathbf{a}^j$), the packing is transformed to the schedule of original PSP according to the problems' equivalence, as was outlined in \cref{sec:packing-section}. For details, see \cite{grus2024icores, hanen2020periodic}. The coordinates of the rectangles determine the start times of the associated tasks. 

\subsubsection{Inferring Ordered List from Schedule}\label{sec:order-of-tasks}
The schedule provided by the CP is used to create an initial ordered list such that its processing by the FFS heuristics leads to a feasible solution (such a solution is always found for 100\% utilized instances, but it is not always found for less utilized instances as shown in \cref{sec:Feasibility_issues}). 

An ordered list that should produce a feasible solution for the \emph{leftmost} heuristic is obtained simply by sorting the tasks by their start time, concatenating the results obtained for all the resources. However, to be used with the \emph{predecessor} method, this relatively arbitrary ordered list may lead to unexpected behavior since it does not respect inter-resource coupling via precedences. If two consecutive tasks of the chain $\task{i}{k},\task{i+1}{k}$ are put in the ordered list one after another, and their independently scheduled start times cause them to collide, the \emph{predecessor} heuristic will schedule the latter task slightly later to account for it. However, this disrupts the schedule obtained by CP (creating holes) and may eventually lead to infeasibility.

We try to resolve this by employing an order-determining CP model for the \emph{predecessor} method. In this model, each task $\task{i}{k}$ of the instance is associated with an integer variable $a^k_i$. Given the start times of the individual tasks established in the previous section, the CP model is defined using:

\begin{align}
    &\mathtt{AllDiff}(\left\{ a^k_i~|~\forall\task{i}{k} \right\}) \\
    & a^{k}_{i} < a^{l}_{j} & \forall m \in \left\{1,\dots,M\right\},~\forall \task{i}{k} \in \tasksmachine{m},\notag\\ & &\forall \task{j}{l} \in \tasksmachine{m}: \first{i}{k} < \first{j}{l} \label{eq:second-cp}\\
    & a^{k}_{i+1} < a^{k}_{i} & \forall k \in \left\{1,\dots,\mu\right\},~\forall i \in \left\{1,\dots,n^k-1\right\}:\notag\\&&\first{i}{k} + \proctime{i}{k} > \first{i+1}{k} \label{eq:cp2b}
\end{align}

All the tasks across all the resources are processed simultaneously. The solver directly determines the ordered list of tasks with the all-different global constraint \cref{eq:second-cp} - the value of $a_i^k$ is the position of $\task{i}{k}$ in the ordered list. Furthermore, when the successor task $\task{i+1}{k}$ would be scheduled later than planned due to the start time of its predecessor (creating a hole), constraint \cref{eq:cp2b} ensures it is processed earlier than its predecessor from the ordered list and the heuristic does not unexpectedly postpone it. 

\subsubsection{Feasibility Issues}
\label{sec:Feasibility_issues}
The aforementioned CP approach consistently finds feasible solutions for most of the 100\%-utilized instances; we observed this during the experimentation reported in the following sections of the paper. Nevertheless, when utilization drops, it may happen that the heuristic cannot transform the provided order into a feasible solution. As results of \cref{sec:exp-feas} show, this happens at worst for 10\% of instances on the most affected instance set. Furthermore, the local search successfully finds feasible solutions for less-utilized instances even without a warm start.

\begin{figure}
\centering
   \begin{subfigure}[b]{0.85\textwidth}
   \includegraphics[width=0.99\textwidth]{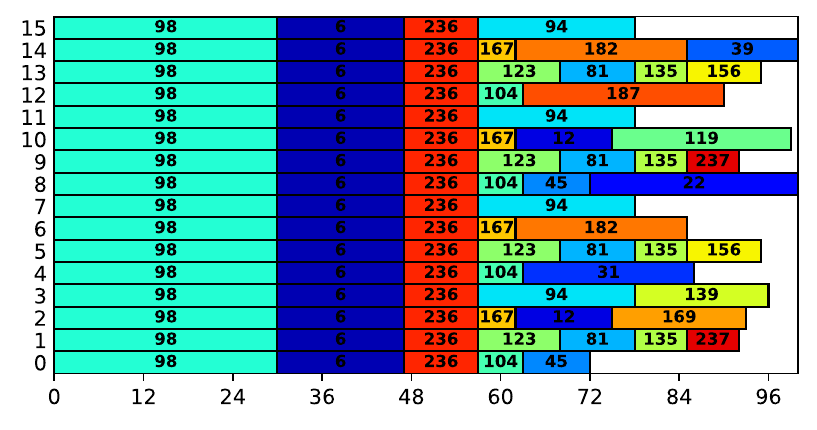}
   \caption{Schedule obtained by CP solver.}
   \label{fig:issue_before} 
\end{subfigure}
\vskip\baselineskip
\begin{subfigure}[b]{0.85\textwidth}
   \includegraphics[width=0.99\textwidth]{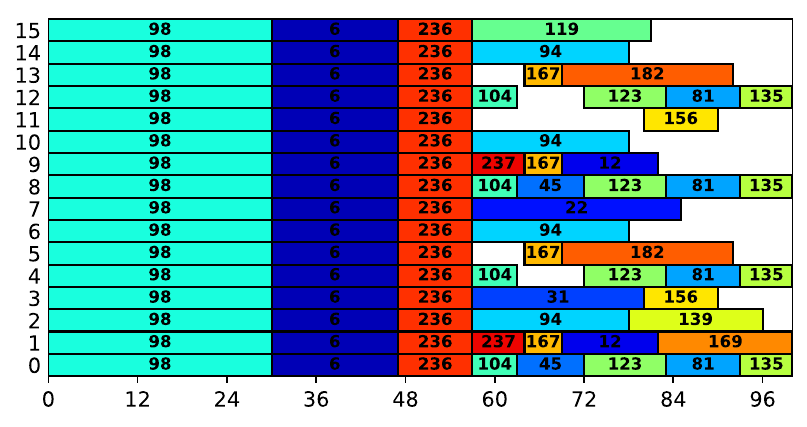}
   \caption{Partial infeasible schedule produced by the FFS heuristic for a given order of tasks deduced from CP schedule.}
   \label{fig:issue_after}
\end{subfigure}
\caption{Schedules documenting the issue with a warm start on less-utilized resources. Notice the gap caused by scheduling task \emph{123} at an incorrect time caused by FFS, leading to infeasibility - task \emph{187} cannot be scheduled.}
\label{fig:issue_of_cp}
\end{figure}

This phenomenon is shown in \cref{fig:issue_of_cp}. In \cref{fig:issue_before}, the schedule of one specific resource produced by the CP warm start is clearly feasible. However, when the FFS heuristic processes tasks row by row, we can see in \cref{fig:issue_after} that it misplaces task \emph{123}; this CP model would put the task into the second row, but the heuristic schedules it in the first one, as it operates in a first fit manner, and there is excessive space available. This creates a gap due to the periodicity of the task. This inefficiency in the schedule eventually leads to the state shown in \cref{fig:issue_after}. Although we were able to schedule all the tasks until that point, once a task \emph{187} (row 12 in \cref{fig:issue_before}) is to be processed, there is no longer enough space to schedule it successfully. 

\subsection{Constraint Programming Matheuristic}\label{sec:matheuristic}

\begin{figure}
        \centering
        \begin{subfigure}[b]{0.49\textwidth}  
            \centering 
            \includegraphics[width=\textwidth]{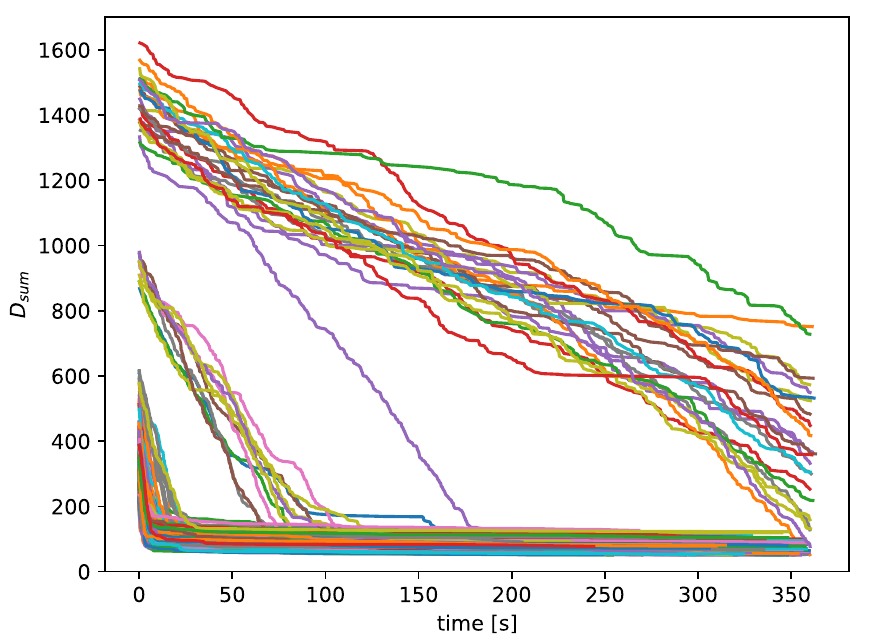}
            \caption{Degeneracy of the best-found solution over time. Each line corresponds to a single instance.\\} 
            \label{fig:mh-curves}
        \end{subfigure}
        \hfill
        \begin{subfigure}[b]{0.49\textwidth}
            \centering
            \includegraphics[width=\textwidth]{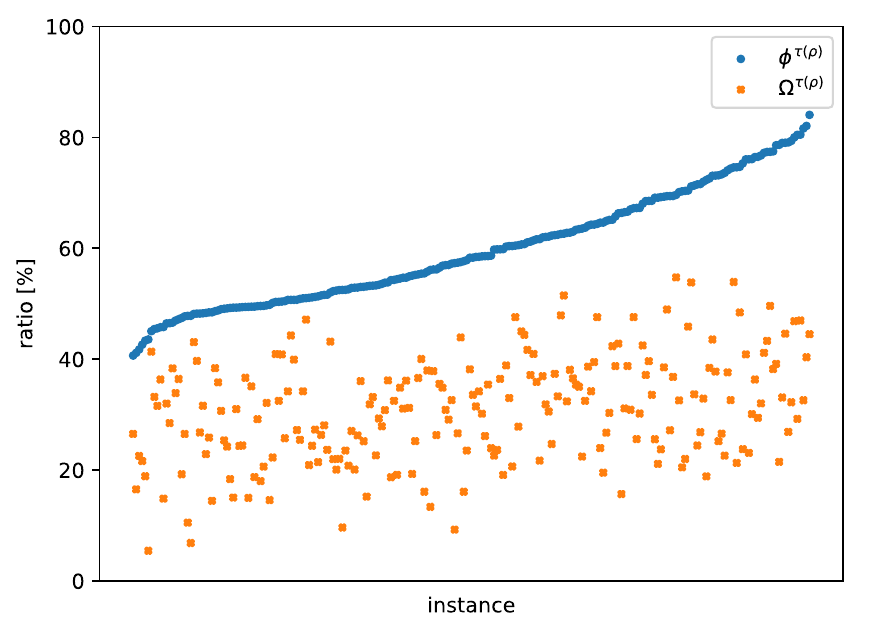}
            \caption{Ratio of degeneracy $\phi^{\per{\rho}}$ and ratio of utilization $\Omega^{\per{\rho}}$ caused by the longest-period chains and tasks.}    
            \label{fig:mh-dots}
        \end{subfigure}
        \caption{Characteristics of instances not solved to optimality by local search.} 
        \label{fig:mh-figures}
\end{figure}

The local search performs generally well, but it sometimes struggles to minimize the degeneracy of highly-utilized instances. We demonstrate this using \cref{fig:mh-figures} (specific results are discussed in \cref{sec:exp-feas,sec:exp-math}). In \cref{fig:mh-curves}, each curve shows the degeneracy of the best-found solution of each instance as the local search progresses. We see that the local search did not converge for several instances in a given time. However, in other cases, the local search rapidly decreased the degeneracy and stagnated with non-zero degeneracy for the rest of the experiment.

We investigated how the degeneracy is distributed within these stagnating instances. We focused on chains (and tasks) with the longest period $\per{\rho}$ only. Blue dots in \cref{fig:mh-dots} show the values of $\phi^{\per{\rho}}$. It is the ratio of degeneracy caused by the longest period chains and the total degeneracy for the final solution of each instance (instances are sorted in ascending order of $\phi^{\per{\rho}}$). We observed that these longest-period chains produce 40\%-80\% of degeneracy, even though they have just a single occurrence in the schedule's hyperperiod. Furthermore, orange dots show values of $\Omega^{\per{\rho}}$, which is the percentage of the utilization allocated to the longest-period tasks on the most utilized resource of an instance. We can see that this value is lower than $\phi^{\per{\rho}}$. This suggests that while the longest-period tasks do not necessarily utilize the instance the most, their chains significantly contribute to the overall degeneracy. The culprit is probably the fact that the ordered list generated by the local search is generally almost rate-monotonic. Since the FFS heuristic processes these tasks later, there are fewer options to schedule them with low degeneracy.

This observation motivated the use of a matheuristic to improve the schedule further. Similarly, as in \cite{della-matheur,greet2020RR}, the focus is a natural way of decomposition: individual periods. If all the tasks with periods shorter than $\per{\rho}$ would be fixed, the longest period tasks could be re-scheduled within the complete hyperperiod $\left[0;\per{\rho} - 1 \right]$. 
Other tasks behave as fixed obstacles. Similarly, we can focus on optimizing chains with arbitrary period $\per{\beta}$. 

 We focus on chains with  $\perjob{k}=\per{\beta}$. Let $\tasksmachine{m}^O$ be a set of tasks with period $\per{\beta}$ on resource $m$. Let $\tasksmachine{m}^F$ be a set of all occurrences of other tasks on resource $m$. To avoid any possible collision, tasks with shorter periods are ``multiplied'' in $\tasksmachine{m}^F$ to account for each of their occurrences within $\per{\beta}$; task with a period $\per{\beta'}$ adds $\per{\beta}/\per{\beta'}$ properly distanced occurrences to $\tasksmachine{m}^F$. Tasks with longer periods are simply projected using the modulo operator. If two occurrences in $\tasksmachine{m}^F$ immediately follow each other, we can merge them into one to simplify the model. These sets are created for all resources. 
 
 As in \cref{sec:structural-properties}, each resource's start time is shifted so that a task with the shortest period starts at time 0, let $o_m$ be such offset. Let $\start{i}{k}$ be the shifted core schedule of the task, as it is in \cref{sec:structural-properties}. Finally, the set of chains with period $\perjob{k} = \per{\beta}$ is denoted as $\jobperset{\per{\beta}}$.

We formulate the following matheuristic CP model. The model uses standard interval variables which tie together start times and processing times and which enable the use of efficient scheduling constraints to handle both the optimized tasks and the fixed occurrences of other tasks. The model tries to re-schedule tasks with period $\per{\beta}$ to decrease the objective function while keeping the other tasks (their occurrences) fixed:

\begin{align}
    & \min ~ \sum_{\job{k} \in \jobperset{\per{\beta}}} (\lceil \SE{k} / \per{\beta}\rceil - 1) && \\
    & \mathtt{NoOverlap}(\tasksmachine{m}^O \cup \tasksmachine{m}^F) && \forall {m} \in \left\{1,\dots,M\right\}\label{eq:no-over} \\
     & \mathtt{StartOf}(\task{i}{k}) = \start{i}{k}&& \forall {m} \in \left\{1,\dots,M\right\},~\forall \task{i}{k} \in \tasksmachine{m}^F\label{eq:fixed-pos}\\
     & \mathtt{StartOf}(\task{i}{k}) \ge 0 && \forall {m} \in \left\{1,\dots,M\right\},~ \forall \task{i}{k} \in \tasksmachine{m}^O\label{eq:var-pos1} \\
     & \mathtt{EndOf}(\task{i}{k}) \le \per{\beta} && \forall {m} \in \left\{1,\dots,M\right\},~ \forall \task{i}{k} \in \tasksmachine{m}^O\label{eq:var-pos2}\\     
     & ps_i^k = (\mathtt{StartOf}(\task{i}{k}) + o_{\pi_i^k }) \mod \per{\beta} && \forall \job{k} \in \jobperset{\per{\beta}},~\forall i \in \left\{1,\dots,n^k\right\}\\
 & pe_i^k = (\mathtt{EndOf}(\task{i}{k}) + o_{\pi_i^k}) \mod \per{\beta} && \forall \job{k} \in \jobperset{\per{\beta}},~\forall i \in \left\{1,\dots,n^k\right\}\\
     & {df}_0^k = 0 && \forall \job{k} \in \jobperset{\per{\beta}}  \\
     & pd_{i+1}^k =  (ps_{i+1}^k - pe_i^k) < 0&& \forall \job{k} \in \jobperset{\per{\beta}},\notag\\&&&\forall i \in \left\{1,\dots,n^k-1\right\}\label{eq:cond-1} \\
     &  {df}_{i+1}^k = ps_{i+1}^k - pe_i^k + \per{\beta}\cdot pd_{i+1}^k  && \forall \job{k} \in \jobperset{\per{\beta}},\notag\\ &&&\forall i \in \left\{1,\dots,n^k-1\right\}\label{eq:cond-2} \\
     & \SE{k} = \sum_{\forall \task{i}{k}\in \job{k}} \proctime{i}{k} +  {df}_{i}^k  && \forall \job{k} \in \jobperset{\per{\beta}}\label{eq:cr-element}\\
    &\task{i}{k}~:~\mathtt{intervalVar} &&\forall {m} \in \left\{1,\dots,M\right\},\notag\\ &&&\forall \task{i}{k} \in \tasksmachine{m}^F\cup \tasksmachine{m}^O\\
    & \SE{k}\in \mathbb{Z} && \forall \job{k} \in \jobperset{\per{\beta}} \\
    &ps_i^k, pe_i^k, df_i^k \in \mathbb{Z} &&\forall \job{k} \in \jobperset{\per{\beta}},~\forall i \in \left\{1,\dots,n^k\right\}\\
    &pd_i^k \in \left\{0,1\right\}&&\forall \job{k} \in \jobperset{\per{\beta}},~\forall i \in \left\{1,\dots,n^k\right\}
\end{align}

\cref{eq:no-over} ensures the non-overlapping of intervals on each resource. \cref{eq:fixed-pos,eq:var-pos1,eq:var-pos2} either fix or limit the intervals of the tasks (their occurrences). Variables $ps_i^k$ and $pe_i^k$ correspond to the start- and completion time of the task once the resource offset is considered. These (with helper binary variables $pd_i^k$) are used to calculate $df_i^k$ -  the consecutive differences between completion- and start times of the tasks in \cref{eq:cond-1,eq:cond-2}. These, together with tasks' processing times, form the chain's start-to-end latency $\SE{k}$ in equation \cref{eq:cr-element}. As in the case of the FFS heuristic, \cref{eq:cond-1,eq:cond-2} regarding the calculation of $df_i^k$ could be modified to account for link delays or cut-through mechanisms. 

The objective function precisely models the degeneracy $\DGs$. Note that both $\DGm$ and $\DGsalpha{\alpha}$ can be modeled in a similar manner. Furthermore, the following constraint \cref{eq:test} is used with sum-based objectives to ensure that the degeneracy of individual chains is not worsened. Let $\DGk{k}$ be the degeneracy of chain $\job{k}$ given the current solution. Then, the newly added constraints are:
\begin{align}
    & \SE{k} \le (\DGk{k} + 1) \cdot \per{\beta} && \forall \job{k} \in \jobperset{\per{\beta}} \label{eq:test}
\end{align}

For large instances, the performance of the CP solver starts to diminish due to the large number of task intervals; the solver cannot reduce the degeneracy of any chain even though the search space is large. Thus, when the size of the problem exceeds a defined bound, not all chains are optimized simultaneously; rather, we optimize only $K$ chains with the largest value of degeneracy (keeping others fixed). With this reduction, the solver can again improve the objective within the provided time. The process is repeated (each time with $K$ most problematic chains) until a time limit is reached.

\section{Experiments}
\label{sec:experiments}
We used C++ and Python 3.10. Experiments were performed on Intel Xeon E5-2690 using a single thread each time. CP Optimizer v22.1 was used as a CP solver.

The experiments were primarily performed on synthetically generated instances inspired by those in \cite{hladik2020complexity}. 
Initially, each resource is split into segments that become tasks of period $\per{1}$ and segments that are recursively split further to allocate longer-period tasks. 
Base period $\per{1}$ was set to 100, 200, or 400, and there were between 3 to 5 distinct periods per instance, with ratios of 2, 3, and 4 between consecutive periods. This ensures that each instance is feasible. Chains are generated afterward, ensuring that the degeneracy of any synthetic instance is zero. Thus, the synthetic instances generated in this manner all admit a feasible solution with an optimal degeneracy of zero, making the comparison of the methods easy.

\begin{figure}
    \centering
    \begin{subfigure}[b]{0.4\textwidth}
        \centering
        \includegraphics[width=\textwidth]{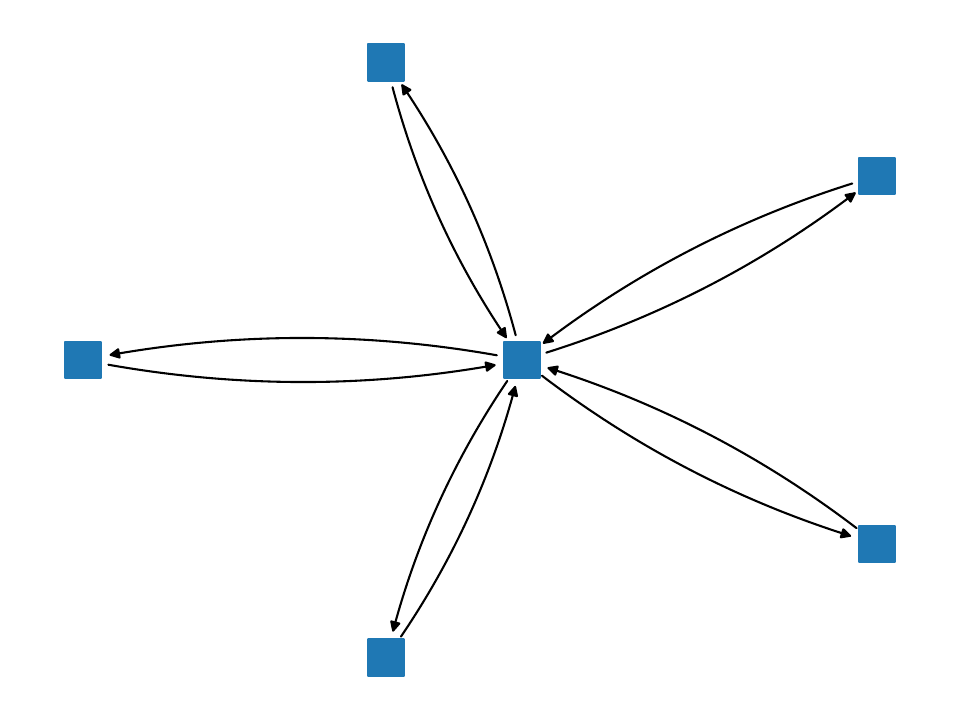}
        \caption{Star topology}    
        \label{fig:star-topology}
    \end{subfigure}
    \hfill
    \begin{subfigure}[b]{0.4\textwidth}  
        \centering 
        \includegraphics[width=\textwidth]{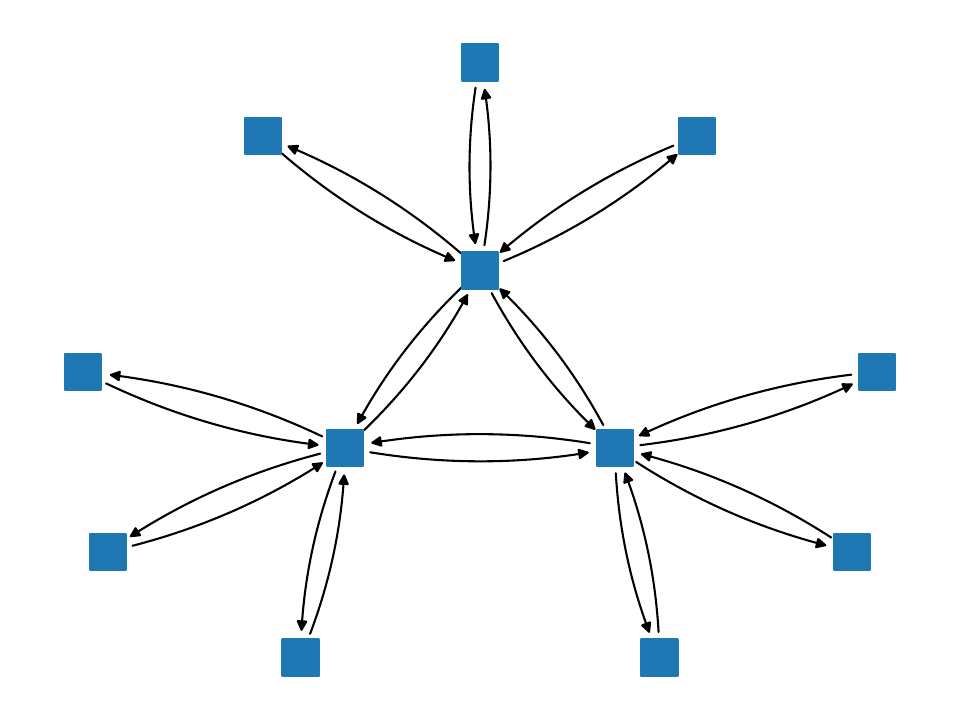}
        \caption{Triangle topology}    
        \label{fig:triangle-topology}
    \end{subfigure}
    \vskip\baselineskip
    \begin{subfigure}[b]{0.4\textwidth}   
        \centering 
        \includegraphics[width=\textwidth]{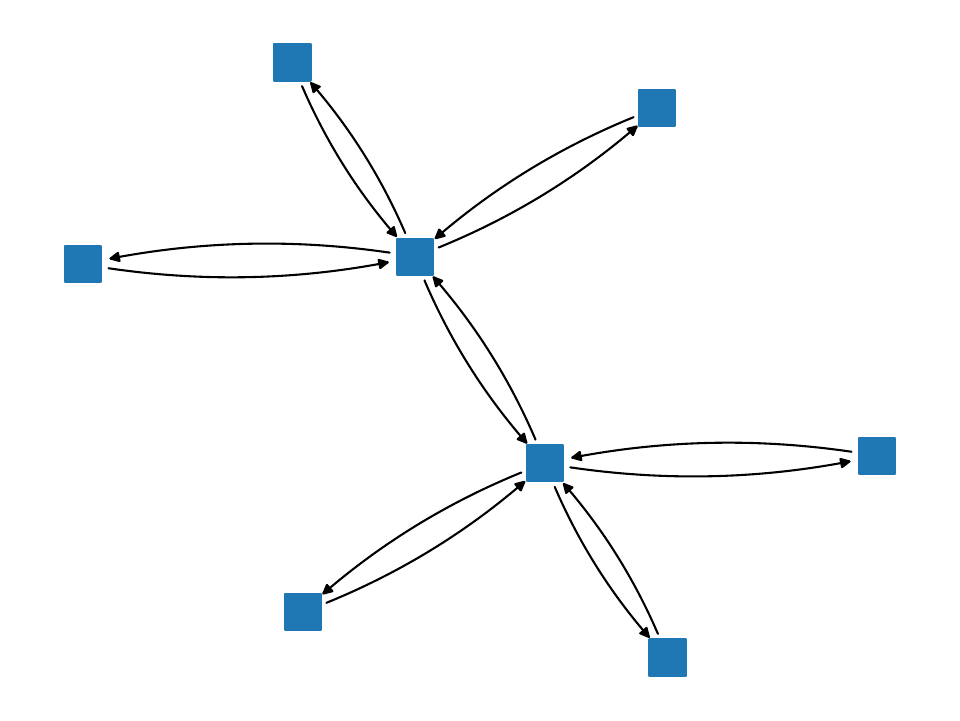}
        \caption{Bridge topology}    
        \label{fig:bridge-topology}
    \end{subfigure}
    \hfill
    \begin{subfigure}[b]{0.4\textwidth}   
        \centering 
        \includegraphics[width=\textwidth]{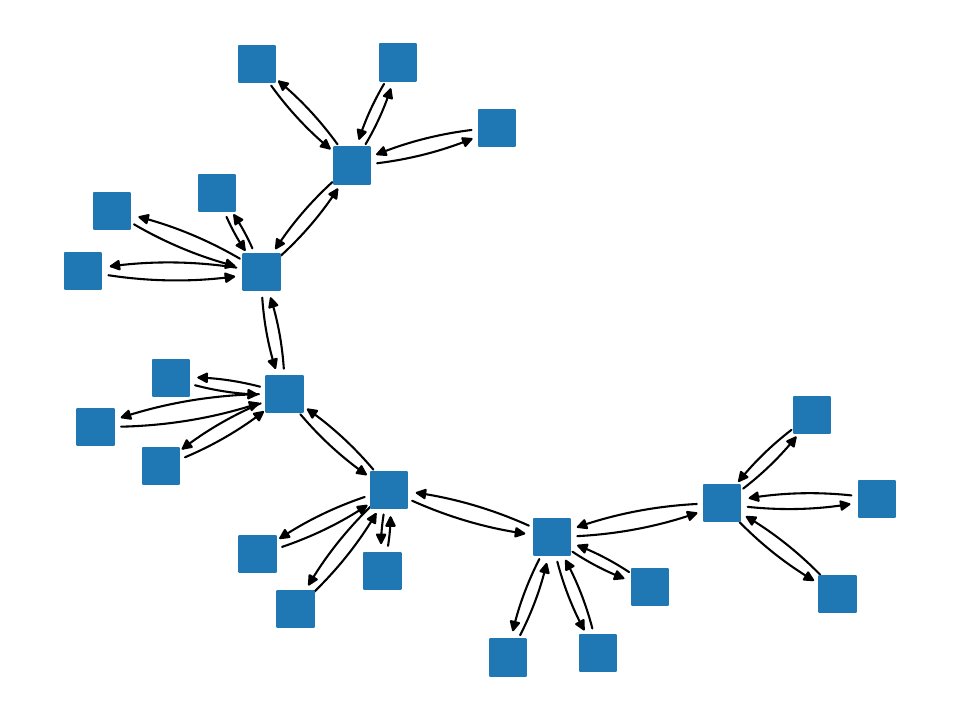}
        \caption{Stubbed line topology}    
        \label{fig:line-topology}
    \end{subfigure}
    \caption{Topologies used for \emph{TOP} (a, b, c) and \emph{LINE} (d) instance sets.}
    \label{fig:topologies}
\end{figure}

Three kinds of instance sets were produced. Set \emph{GEN-x} contains the instances for the general case of PSP with 5 or 10 resources. Chains can pass from one arbitrary resource to another, even returning to the original one; the support graph of such an instance can be a complete graph if the instance's chains induce it. This corresponds to the most general case of PSP. Furthermore, the processing times of tasks of a single chain can differ. \emph{x} refers to  $\min_m U_m$; the minimum utilization of resources. 

Sets \emph{TOP} and \emph{LINE} were generated using four different underlying topologies: star, triangle, bridge, and stubbed line (used topologies are shown in \cref{fig:topologies}). \emph{LINE} contains instances using the stubbed line topology; the rest is put into the set \emph{TOP}. This was done so the results on more complex stubbed-line instances with longer chains would not be skewed by the instances generated using smaller topologies. Processing time is the same for all tasks of each chain. The links are full-duplex, and messages are sent between endpoints. During generation, two random endpoint network nodes were sampled, and a chain was created between them if possible. This chain sampling was repeated until no more chains with zero degeneracy could be scheduled. Furthermore, there are also tasks being sent between two neighboring nodes only (essentially forming a chain of length 1) to increase the utilization. These additional tasks were omitted to obtain less-utilized sets  \emph{TOP-LESS}, \emph{LINE-LESS}).

Finally, \emph{THEORY} set is used to discuss the complexity results; it consists of acyclic instances with a bottleneck resource. Specifically, these instances correspond to scattering cases with line and tree support graphs (see \cref{sec:line-tree-top}). The parameters of the instance sets are presented in \cref{tab:instance-properties}.

 \begin{table}[htbp]
 \centering
 \adjustbox{max width=0.8\textwidth}{%
\begin{tabular}{lcccccc}\toprule
 & \# instances & \# resources & avg \# tasks & avg \# chains & avg chain length & avg $\min_m U_m$ \\
\midrule
\emph{GEN-1} & 1194 & 5-10 & 2193 & 254 & 8.4 & 1.00 \\
\emph{GEN-0.98}  & 1194  & 5-10& 1786 & 208 & 8.3 & 0.98\\
\emph{GEN-0.96} & 1194  & 5-10& 1558 & 183 & 8.2 & 0.96\\
\emph{GEN-0.94}   &  1194  & 5-10& 1374 & 163 & 8.1 & 0.94\\
\emph{GEN-0.9}& 1194  & 5-10& 1093 & 131 & 7.9 & 0.90 \\
\emph{LINE}& 796 & 46 & 7634 & 2247 & 2.7 & 1.00\\
\emph{LINE-LESS}& 796 & 46 & 6155 & 2246 & 2.7 & 0.71\\
\emph{TOP}& 2388 & 10-24 & 2487 & 792 & 2.3 & 1.00 \\
\emph{TOP-LESS}& 2388 & 10-24 & 1929 & 792 & 2.35 & 0.73 \\
\emph{THEORY}& 320 & 10-77 & 1102 & 96 & 11.2 & 0.66\\
\bottomrule
\end{tabular}}
 \caption{Parameters of synthetically generated instance sets.}
 \label{tab:instance-properties}
\end{table}

\subsection{Finding Feasible Solutions}\label{sec:exp-feas}
First, we studied the effect of a warm start and the ability of the local search to find feasible and high-quality solutions. We ended with a strategy that uses the CP to warm-start the heuristic for instances with utilization close to 100\%; otherwise, it relies on the local search itself. This strategy provided a good trade-off between the feasibility success rate and the value of the objective function.

The experiment was performed on \emph{TOP}, \emph{LINE}, and \emph{GEN} instances with a time limit of 6 minutes. In \cref{tab:heur-warm}, we report the percentage of instances where the feasible solution was found, the median (and standard deviation) of degeneracy $\DGs$, and a percentage of instances with 0-$\DGs$ solutions (when the final solution has degeneracy equal to 0). Note that the values of $\DGs$ are reported only for a subset of instances for which the method found any solution at all. Thus, it is necessary to compare both the ability to find feasible solutions and the ability to optimize them efficiently when analyzing the results so that a practically inapplicable method that is efficient in a few easy instances does not skew the conclusions. Finally, we also show the median of $\DGs^{chains}$, which is the degeneracy $\DGs$ divided by the number of chains. This metric describes the delay of a mean chain; however, different lengths of chains and their number may skew the resulting value.

The results are reported for both \emph{leftmost} and \emph{predecessor} heuristic methods, either being warm-started with \emph{CP} or starting from the default ordered list. The last method \emph{predecessor-flow} follows \cref{fig:method-diagram}). It starts a local search from the default ordered list, and if it does not find a feasible solution within 15 seconds (giving local search sufficient time to initialize)
, it runs a CP warm start instead (\cref{sec:warmstart}).

 \begin{table}[htbp]
 \centering
 \adjustbox{max width=0.7\textwidth}{%
\begin{tabular}{lcccccccc}\toprule
& \multicolumn{4}{c}{\emph{leftmost}} & \multicolumn{4}{c}{\emph{leftmost-CP}}\\
\cmidrule(lr){2-5}\cmidrule(lr){6-9}
  & \% feasible & $\DGs$ &$\DGs^{chains}$ & \% 0-$\DGs$
           & \% feasible & $\DGs$ &$\DGs^{chains}$ & \% 0-$\DGs$ \\

\midrule
\emph{GEN-1}& 21.4 & \textbf{378.5 (538.2)} & \textbf{2.1 (0.6)} & \textbf{0.0}& \textbf{100.0} & 621.5 (624.3) & 3.0 (0.4) & \textbf{0.0}\\
\emph{GEN-0.98}& \textbf{100.0} & 220.0 (328.4) & 1.3 (0.4) & 0.0& 89.4 & 467.0 (511.6) & 2.6 (0.6) & 0.0\\
\emph{GEN-0.96}& \textbf{100.0} & 198.0 (236.8) & 1.3 (0.3) & 0.0& 98.3 & 369.5 (430.7) & 2.5 (0.6) & 0.0\\
\emph{GEN-0.94}& \textbf{100.0} & 181.5 (204.3) & 1.3 (0.3) & 0.0& 99.1 & 321.0 (372.8) & 2.3 (0.6) & 0.0\\
\emph{GEN-0.9}& \textbf{100.0} & 151.0 (155.8) & 1.3 (0.4) & 0.0& \textbf{100.0} & 236.0 (285.3) & 2.1 (0.7) & 0.0\\
\emph{LINE}& 69.1 & 889.5 (326.0) & 0.4 (0.0) & \textbf{0.0}& \textbf{100.0} & 775.0 (268.4) & 0.4 (0.0) & \textbf{0.0}\\
\emph{LINE-LESS}& \textbf{100.0} & 721.0 (334.3) & 0.4 (0.0) & 0.0& 85.3 & 720.0 (272.9) & 0.3 (0.0) & 0.0\\
\emph{TOP}& 69.0 & 75.0 (165.1) & 0.1 (0.1) & 2.0& \textbf{99.9} & 110.5 (166.2) & 0.2 (0.1) & 6.7\\
\emph{TOP-LESS}& \textbf{100.0} & 48.0 (128.1) & 0.1 (0.1) & 3.9& 92.0 & 55.0 (143.8) & 0.1 (0.1) & 29.3\\
     \bottomrule
\end{tabular}
}
\bigskip

 \adjustbox{max width=\textwidth}{%
\begin{tabular}{lcccccccccccc}\toprule
& \multicolumn{4}{c}{\emph{predecessor}} & \multicolumn{4}{c}{\emph{predecessor-CP}} & \multicolumn{4}{c}{\emph{predecessor-flow}}\\
\cmidrule(lr){2-5}\cmidrule(lr){6-9}\cmidrule(lr){10-13}
 & \% feasible & $\DGs$ &$\DGs^{chains}$ & \% 0-$\DGs$
  & \% feasible & $\DGs$ &$\DGs^{chains}$ & \% 0-$\DGs$
           & \% feasible & $\DGs$ &$\DGs^{chains}$ & \% 0-$\DGs$ \\
\midrule
\emph{GEN-1}& 0.1 & 651.0 (0.0) & 2.3 (0.0) & \textbf{0.0}& 85.9 & 610.0 (626.8) & 3.0 (0.4) & \textbf{0.0}& 85.8 & 611.0 (621.3) & 3.0 (0.4) & \textbf{0.0}\\
\emph{GEN-0.98}& \textbf{100.0} & 90.0 (181.2) & \textbf{0.6 (0.4)} & 3.0& 73.3 & 256.0 (286.0) & 1.5 (0.5) & 0.0& 99.0 & \textbf{87.5 (201.7)} & 0.6 (0.5) & \textbf{3.2}\\
\emph{GEN-0.96}& \textbf{100.0} & \textbf{11.0 (69.5)} & \textbf{0.1 (0.2)} & \textbf{26.0}& 83.7 & 106.0 (122.3) & 0.8 (0.4) & 0.6& \textbf{100.0} & \textbf{11.0 (86.0)} & 0.1 (0.2) & 25.0\\
\emph{GEN-0.94}& \textbf{100.0} & \textbf{0.0 (11.1)} & \textbf{0.0 (0.1)} & \textbf{51.8}& 86.4 & 46.0 (58.8) & 0.4 (0.3) & 2.4& \textbf{100.0} & \textbf{0.0 (21.3)} & \textbf{0.0 (0.1)} & 50.6\\
\emph{GEN-0.9}& \textbf{100.0} & \textbf{0.0 (2.6)} & \textbf{0.0 (0.0)} & 80.2& 87.4 & 12.0 (28.5) & 0.1 (0.2) & 13.5& \textbf{100.0} & \textbf{0.0 (2.6)} & \textbf{0.0 (0.0)} & \textbf{81.2}\\
\emph{LINE}& 23.7 & \textbf{243.0 (152.4)} & \textbf{0.1 (0.0)} & \textbf{0.0}& 82.8 & 719.0 (231.0) & 0.4 (0.0) & \textbf{0.0}& 80.8 & 684.0 (261.8) & 0.4 (0.1) & \textbf{0.0}\\
\emph{LINE-LESS}& \textbf{100.0} & \textbf{0.0 (64.5)} & \textbf{0.0 (0.0)} & \textbf{84.7}& 75.8 & 147.0 (284.3) & 0.1 (0.1) & 2.4& 99.7 & \textbf{0.0 (78.2)} & \textbf{0.0 (0.0)} & 81.3\\
\emph{TOP}& 64.2 & \textbf{2.0 (27.3)} & \textbf{0.0 (0.0)} & \textbf{25.0}& 98.0 & 101.0 (148.1) & 0.2 (0.1) & 2.8& 99.0 & 9.0 (135.3) & 0.0 (0.1) & 22.3\\
\emph{TOP-LESS}& \textbf{100.0} & \textbf{0.0 (0.1)} & \textbf{0.0 (0.0)} & 99.7& 91.2 & \textbf{0.0 (35.4)} & \textbf{0.0 (0.0)} & 70.9& \textbf{100.0} & \textbf{0.0 (0.0)} & \textbf{0.0 (0.0)} & \textbf{99.8}\\
     \bottomrule
\end{tabular}
}
 \caption{Results of non-warm-started and warm-started local search methods.}
 \label{tab:heur-warm}
\end{table}

Firstly, the \emph{leftmost}-based methods are exceptional at finding feasible solutions, especially when the CP warm start is used. However, since they do not consider chains during scheduling, their degeneracy values are larger, even more so once the utilization of instances decreases. Therefore, they are not suitable in our TSN-like scenario. Notice, however, that the results of \emph{leftmost-CP} for \emph{GEN-1} instance set report almost the same values of degeneracy as the other methods.

We can see that the \emph{predecessor} FFS heuristic achieves great results if the utilization is below  98\%. It finds a feasible solution and achieves low values of $\DGs$, as well as $\DGs^{chains}$. In the case of \emph{predecessor-CP}, bin-packing CP solver finds a feasible schedule for a warm start in most cases 
, but it may not produce an ordered list acceptable by the FFS heuristic (see \cref{sec:order-of-tasks}). Even when the ordered list succeeds, final values of $\DGs$ seem to be worse than in the case of \emph{predecessor} results. The reason may be that the initial ordered list lacks rate-monotonicity, making it harder for local search to improve it. However, CP warm start helps to solve the most utilized instances among our sets \emph{LINE} and \emph{GEN-1}.

The results of the hybrid \emph{predecessor-flow} approach are provided in the last column of \cref{tab:heur-warm}. We can see that even this simple strategy, applying CP warm start when needed (as outlined in \cref{fig:method-diagram}), maintains high success rates and low degeneracy values. Thus, we use this approach in the rest of the paper.

\subsection{Benefit of Matheuristic}
\label{sec:exp-math}
In this section, we discuss the benefits of the matheuristic CP,  which is used to improve the degeneracy of the feasible schedule found by local search. Firstly, we determined that for a larger number of tasks, the best course of action is to optimize only several chains of the same period at once. Empirically, we set the number of non-fixed tasks of the model to at most 500 and the time limit for each optimization iteration to 10 seconds. 

We compared two ways how to use matheuristic: (i) \emph{fixed matheuristic}, when the matheuristic was initiated after five minutes of local search; and (ii) \emph{dynamic matheuristic}, when the matheuristic starts once the local search does not improve the solution for 60 seconds. We determined this dynamic threshold from the behavior of local search (its convergence) on a subset of instances (see \cref{fig:mh-curves}), so the matheuristic would not start prematurely. The local search with no matheuristic is used as a baseline. In all cases, the computation time was six minutes.

In \cref{tab:matheur}, we compare the degeneracy obtained while using fixed or dynamic matheuristic ($\DGs^{\mathrm{MH}}$) and the degeneracy obtained by the local search alone ($\DGs$). We report median (and standard deviation) values of $\DGs^{\mathrm{MH}}$ and differences $\Delta\DGs=\DGs^{\mathrm{MH}} - \DGs$. Thus, a negative value of $\Delta\DGs$ means that, on average, using matheuristic was beneficial.

 \begin{table}[htbp]
 \centering
 \adjustbox{max width=0.7\textwidth}{%
\begin{tabular}{lcccc}\toprule
& \multicolumn{2}{c}{\emph{fixed matheuristic}} & \multicolumn{2}{c}{\emph{dynamic matheuristic}} \\
\cmidrule(lr){2-3}\cmidrule(lr){4-5}
    & $\DGs^{\mathrm{MH}}$       & $\Delta\DGs$ & $\DGs^{\mathrm{MH}}$ & $\Delta\DGs$\\           
\midrule

\emph{GEN-1} & \textbf{458.5 (542.2)} & \textbf{-139.0 (93.7)} & 546.5 (518.4) & -26.0 (169.0)\\
\emph{GEN-0.98} &73.5 (181.4) & \textbf{-12.0 (35.4)} & \textbf{69.5 (178.1)} & -9.0 (16.0) \\
\emph{GEN-0.96} & \textbf{11.0 (83.8)} & \textbf{0.0 (36.8)} & 11.5 (68.6) & \textbf{0.0 (7.7)} \\
\emph{GEN-0.94} & \textbf{0.0 (14.7)} & \textbf{0.0 (11.3)} & 1.0 (10.3) & \textbf{0.0 (1.8)} \\
\emph{GEN-0.9}& \textbf{0.0 (2.5)} & \textbf{0.0 (0.3)} & \textbf{0.0 (2.8)} &  \textbf{0.0 (0.5)} \\
\emph{LINE} & \textbf{552.0 (237.7)} & \textbf{-156.0 (65.3)} & 642.5 (277.9) & -28.0 (59.0)\\
\emph{LINE-LESS}& \textbf{0.0 (61.3)}& \textbf{0.0 (17.3)} &  \textbf{0.0 (71.6)} & \textbf{0.0 (43.3)}\\
\emph{TOP} & 34.0 (78.9) & \textbf{-60.0 (89.9)} & \textbf{7.0 (111.5)} & -3.0 (54.0)\\
\emph{TOP-LESS}& \textbf{0.0 (0.0)} & \textbf{0.0 (0.0)} &  \textbf{0.0 (0.0)} & \textbf{0.0 (0.0)}\\
\bottomrule

\end{tabular}}
 \caption{Effect of fixed and dynamic matheuristic on values of $\DGs$.}
 \label{tab:matheur}
\end{table}

\begin{figure}
    \centering
    \includegraphics[width=0.5\linewidth]{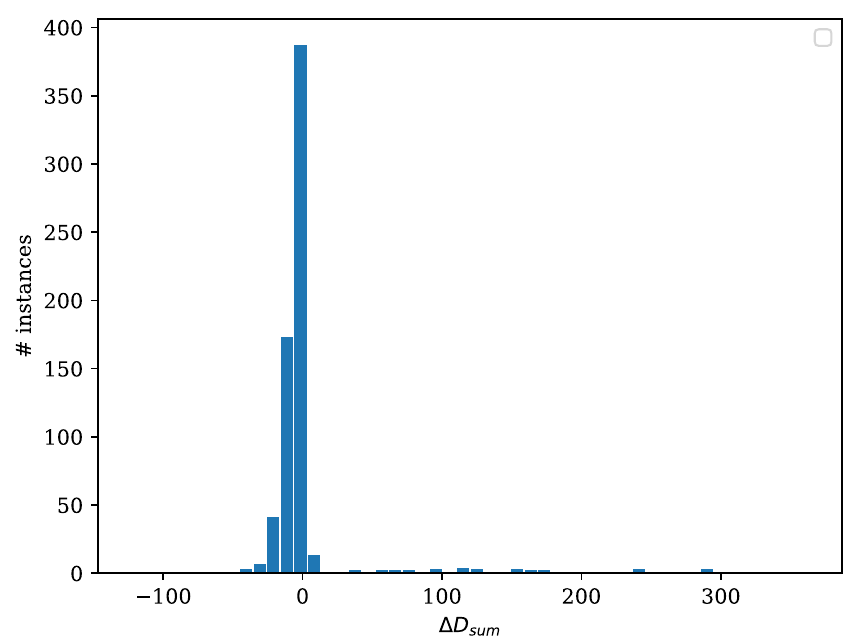}
    \caption{Histogram of values of $\Delta\DGs$ for instance set \emph{GEN-0.96} and \emph{fixed matheuristic}.}
    \label{fig:lng96_without_zero}
\end{figure}

Especially for the maximally-utilized instances, the benefit of the matheuristic is significant - the newly obtained values of $\DGs^{\mathrm{MH}}$ are significantly lower than the previous values $\DGs$. However, for the \emph{fixed matheuristic}, its benefit drops significantly and even diminishes for some instances - the local search probably has not yet converged, and running it further could yield a better solution. This can be seen in \cref{fig:lng96_without_zero}. The histogram of values $\Delta\DGs$ is shown for instances \emph{GEN-0.96} (for clarity, only those with $\Delta\DGs \neq 0$ are presented). We can see that the majority of the differences are negative, meaning the matheuristic was beneficial. However, a few instances with very large positive values of $\Delta\DGs$ skew the average of the difference values to $2.9$ - using matheuristic would be detrimental.

While the dynamic setting does not produce as ground-breaking beneficial examples as the fixed one for the most highly-utilized instances, we observed that its results are more consistent across the instance sets; e.g., on \emph{GEN-0.98}, the most detrimental effect of \emph{dynamic matheuristic} was $\Delta\DGs=+19.0$. This is a much better overall result than the worst case reported for the \emph{fixed matheuristic} ($+253$) for the same instance set. Therefore, we used the \emph{dynamic matheuristic} in the rest of the paper.

\subsection{Time Complexity of Special Cases}\label{sec:exp-line}
In this section, we looked into scheduling for acyclic instances with a bottleneck resource, specifically instances with line and out-tree support graphs. We illustrate how well the \emph{offset schedule} approach of \cref{sec:line-tree-top} outperforms our proposed methodology in these special cases.

To do this, we utilized the set \emph{THEORY} with 320 instances, which upholds the properties outlined \cref{sec:line-tree-top}. We compare the best results obtained by our proposed flow of \cref{sec:methods} with the \emph{offset schedule} approach described in \cref{sec:offset-schedule}: we attempt to find a feasible schedule using the CP model of \cref{sec:warmstart} for the first common resource among the chains, and then we copy the schedule to other resources with appropriate offsets.

 \begin{table}[htbp]
 \centering
 \adjustbox{max width=0.9\textwidth}{%
\begin{tabular}{lcccccccc}\toprule
& \multicolumn{4}{c}{\emph{predecessor-flow}} & \multicolumn{4}{c}{\emph{offset schedule}}\\
\cmidrule(lr){2-5}\cmidrule(lr){6-9}

& \% feasible & $\DGs$ &$\DGs^{chains}$ & \% 0-$\DGs$
      & \% feasible & $\DGs$ &$\DGs^{chains}$ & \% 0-$\DGs$\\

\midrule
\emph{THEORY} & 77.8 & 114.0 (335.2) & 2.8 (2.5) &0.0 & 90.1 & 0.0 (0.0) & 0.0 (0.0) &90.1 \\
\bottomrule
\end{tabular}}
 \caption{Comparison of feasible solutions found and final degeneracy obtained by \emph{predecessor-flow} and \emph{offset schedule} on \emph{THEORY} instances.}
 \label{tab:theory-data}
\end{table}

In \cref{tab:theory-data}, we can see that the \emph{offset schedule} wins in this special case, finding more feasible solutions, each with zero degeneracy. The reason is that it only needs to find a feasible solution for a single resource. Then, it immediately obtains a solution with optimal degeneracy. This highlights the crucial role of problem analysis, which can massively simplify PSP. 

\subsection{Different Objectives}
Up until now, we have focused mostly on degeneracy $\DGs$ as the objective. In this section, we briefly study the differences in results when we optimize $\DGsalpha{0.75}$ or $\DGm$ instead. We ran the local search for instances of set \emph{GEN-0.98} with the two aforementioned objectives and compared the results with the baseline results, optimizing $\DGs$.

 \begin{table}[htbp]
 \centering
 \adjustbox{max width=0.6\textwidth}{%
\begin{tabular}{lccc}\toprule
& \multicolumn{3}{c}{objective} \\
\cmidrule(lr){2-4}
   metric    & $\DGs$ & $\DGsalpha{0.75}$ & $\DGm$\\           
\midrule
 $D_{sum}$ & 90.0 (181.1)& 110.0 (228.5)& 124.5 (259.2)\\
$D_{sum}(0.75)$ & 191.0 (280.3)& 191.5 (340.6)& 251.0 (357.3)\\
$D_{\max}$ & 5.0 (1.7)& 5.0 (1.4)& 5.0 (1.5) \\
\bottomrule
\end{tabular}}
 \caption{Average values of metrics given the objective. Results reported for set \emph{GEN-0.98}.}
 \label{tab:crit-table}
\end{table}

\begin{figure}
\centering
\begin{subfigure}[b]{.6\linewidth}
\includegraphics[width=\linewidth]{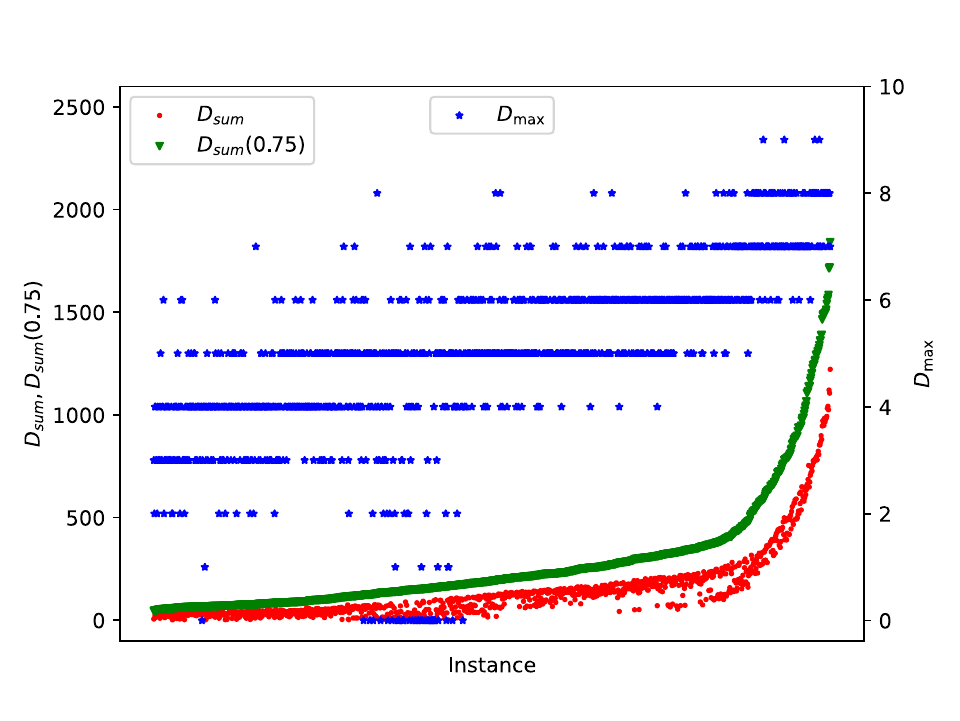}
\caption{Minimizing $\DGs$}\label{fig:crit-deg}
\end{subfigure}

\begin{subfigure}[b]{.6\linewidth}
\includegraphics[width=\linewidth]{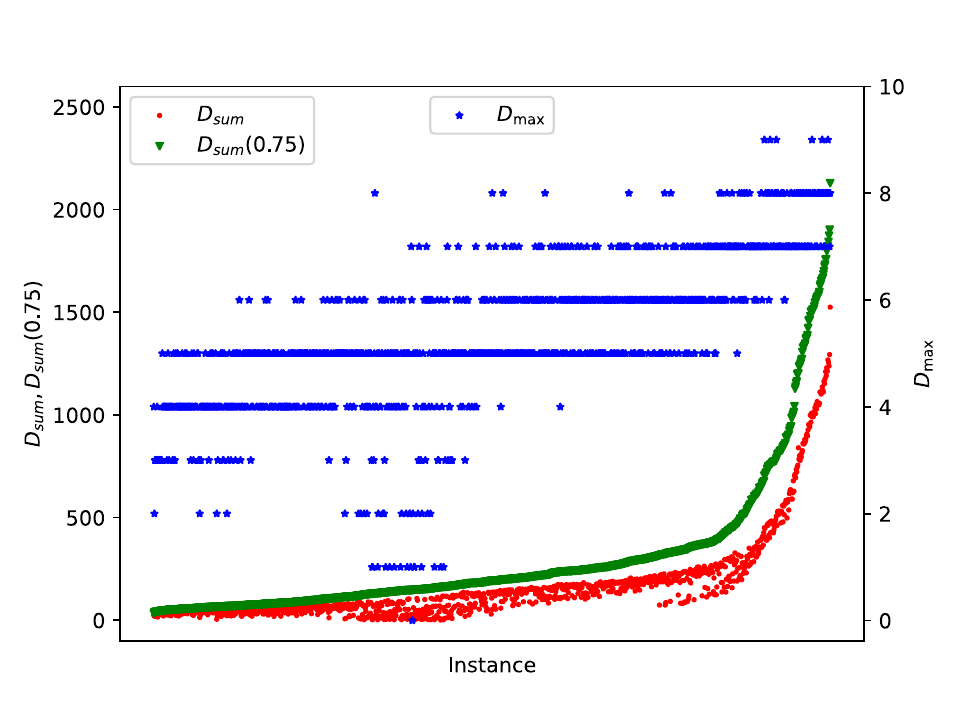}
\caption{Minimizing $\DGsalpha{0.75}$}\label{fig:crit-alpha}
\end{subfigure}

\begin{subfigure}[b]{.6\linewidth}
\includegraphics[width=\linewidth]{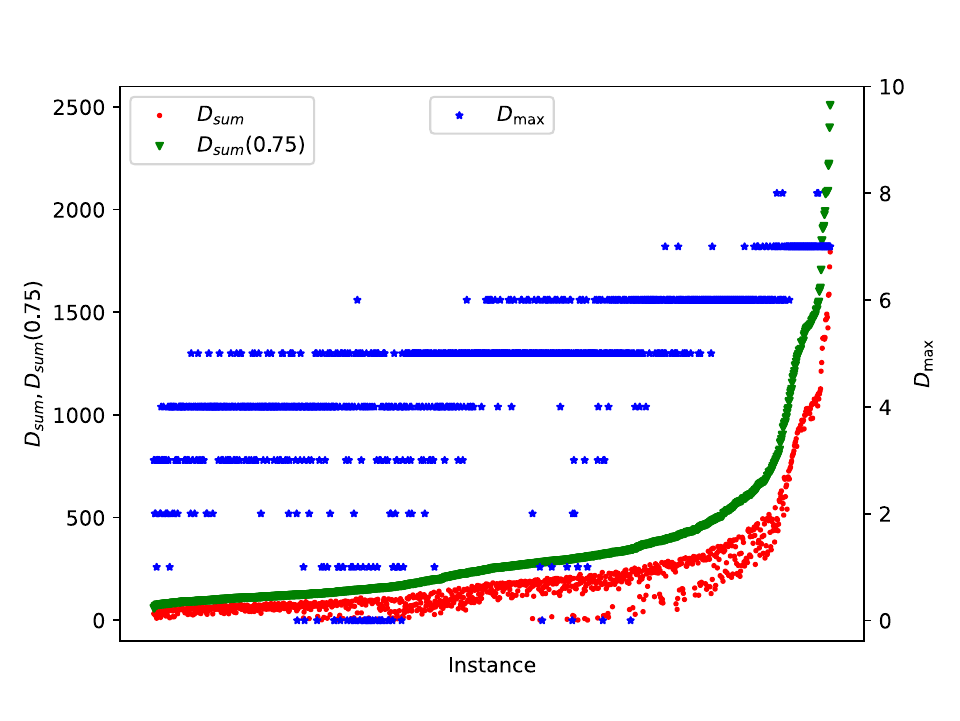}
\caption{Minimizing $\DGm$}\label{fig:crit-max}
\end{subfigure}
\caption{Values of metrics for the instances of \emph{GEN-0.98} for different objectives. Instances were sorted in ascending order of $\DGs$.}
\label{fig:criteria}
\end{figure}

Results per instance are shown in \cref{fig:criteria}. In each figure, three vertically aligned markers show the values of the metrics for a single instance. \cref{fig:crit-deg} shows the values of metrics when $\DGs$ was minimized, \cref{fig:crit-alpha} corresponds to the minimization of $\DGsalpha{0.75}$, and \cref{fig:crit-max} to the minimization of $\DGm$.

From all the figures, it is clear that values of $\DGsalpha{0.75}$ are highly correlated with the original degeneracy in all experiments, even though the used instances do not necessarily admit solution with $\DGsalpha{0.75}=0$ (but they do admit solution with $\DGs=0$). The results obtained by minimizing $\DGs$ in \cref{fig:crit-deg} and $\DGsalpha{0.75}$ in \cref{fig:crit-alpha} are rather similar. However, when $\DGm$ is used as an objective in \cref{fig:crit-max}, the worst reported value of $\DGm$ among the instances is 8 instead of 9. 

Median and standard deviations of the metrics are shown in \cref{tab:crit-table}. There, the difference between optimizing $\DGs$ and $\DGsalpha{0.75}$ is clear. Interestingly, direct optimization of $\DGsalpha{0.75}$ led to an overall slightly worse solution concerning all the metrics than when we used the $\DGs$ objective, even though the space of the feasible solutions is the same. One reason for this is the local search procedure; the initial phase, where the chains are reordered, ends after the first failure to improve the objective function. This happens earlier in the case of the $\DGsalpha{0.75}$. We believe there might be a connection between the worse performance with stricter $\DGsalpha{\alpha}$ objective and the difficulty of finding feasible solutions in a TSN setting with release times and deadlines.
 
\subsection{Real-Life Inspired Instances}
Finally, we performed several experiments on real-life inspired instances.

\subsubsection{Production Line}
Firstly, we used \emph{production-line} instances from \cite{vlk2022largescale}. In that paper, three types of instances were generated, one of which used harmonic periods. The original paper handled problems with more complex instances than we do, including frame-isolation constraints, queues, release times, and deadlines, and reported low success rates once the maximum utilization among the resources rose above 50\%. These additional constraints are omitted in this paper.

 \begin{table}[htbp]
 \centering
 \adjustbox{max width=0.6\textwidth}{%
\begin{tabular}{lcccc}\toprule
           & \% feasible & avg $\DGs$ & avg $\DGs^{chains}$ & \% 0-$\DGs$  \\    
\midrule
\emph{production-line} & 100.0 & 530.9 $\pm$ 1332.5 & 0.1 $\pm$ 0.1& 80.5\\
\bottomrule
\end{tabular}}
 \caption{Feasibility and degeneracy values reported for \emph{production-line} dataset (3000 instances) optimized with \emph{predecessor-flow}.}
 \label{tab:production-line}
\end{table}

In \cref{tab:production-line}, we present the results across the 3000 mentioned instances, obtained after six minutes. We report average values and standard deviations of the degeneracy metrics since their medians were 0. For most instances, we find a 0-$\DGs$ solution. For the rest of the instances, the degeneracy was high. However, in these cases, the number of tasks was, on average, 30000, and the number of chains was 2700. The average $\DGm$ was equal to $6.2$, i.e., the most delayed chain reported degeneracy $<7$. Altogether, better results could be achieved by providing the solver more time, as optimization has not yet converged in these cases. However, it would probably be worth investigating more aggressive neighborhood operators or decomposition techniques for these large-scale instances. Note that CP warm start could still determine feasible schedules even for these large instances.

\subsubsection{Avionic}
The second real-life set is the \emph{avionic} instances, also from \cite{vlk2022largescale}. Chains of the five industrial instances from the avionic domain were scheduled across 70-92 resources, with a maximum utilization of less than 38\%. Thus, we also created a second set of instances in which the periods were halved, doubling the utilization. However, for all 10 instances, our approach found feasible 0-$\DGs$ solutions within two minutes (given the constraints we study in this paper).

\subsubsection{Automotive}
\begin{figure}
        \centering
        \begin{subfigure}[b]{0.49\textwidth}  
            \centering 
            \includegraphics[width=\textwidth]{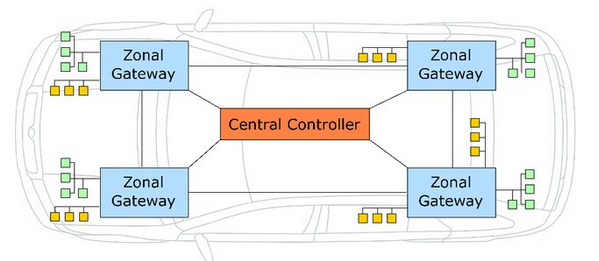}
            \caption{Example of zonal architecture \cite{zonal}.\linebreak} 
            \label{fig:zonal}
        \end{subfigure}
        \hfill
        \begin{subfigure}[b]{0.35\textwidth}
            \centering
            \includegraphics[width=\textwidth]{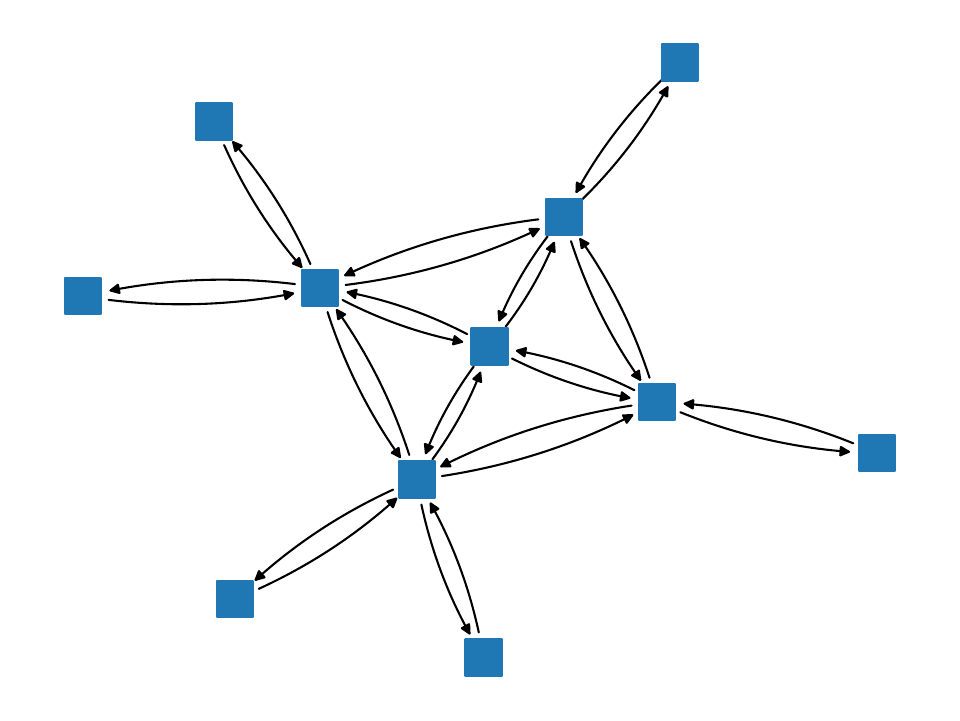}
            \caption{Zonal-architecture inspired network used in the experiment.}    
            \label{fig:zonal-net}
        \end{subfigure}
        \caption{Example of zonal architecture and used network.} 
        \label{fig:zonal-figs}
\end{figure}

Finally, we generated a set of \emph{automotive} instances following a zonal architecture in modern cars. An example of such architecture is shown in \cref{fig:zonal}, which we used as an inspiration for our network in \cref{fig:zonal-net}. The central controller used 5Gb/s communication speed, the encompassing ring used 2.5Gb/s, and the rest of the connections used 1 Gb/s. We generated messages between leaf and central nodes with between 72 to 1542 bytes, as is used in Ethernet frames, and determined the processing times of the jobs based on both the message size and the resource's speed. The messages' periods were 10 ms, 50 ms, 250 ms, and 500 ms.

We generated 20 such instances with an average maximum utilization of $\utilization{m}=0.9$ and 300000+ tasks each. However, due to the maximum length of the Ethernet messages, their processing times were short in contrast to the minimum period. Thus, our approach found feasible solutions with $\DGm \le 2$ within 6 minutes, even for these large-scale instances. Even with just a few iterations, the local search produced a schedule with satisfactory characteristics. 
\section{Conclusion}
\label{sec:conclusion}

In this paper, we presented complexity results related to the periodic scheduling problem with harmonic periods and dedicated resources. Crucially, we have shown that for acyclic cases with a bottleneck resource, the offset schedule approach can be used to find a high-quality solution efficiently. 

We proposed a warm start approach, which successfully helps the local search optimize instances with high utilization. Furthermore, we described a CP matheuristic, which leverages CP scheduling capabilities by optimizing the schedule while targeting one period at a time. 

Several interesting conclusions can be drawn from our results. Even though scheduling periodic tasks is the principal component of TSN, we achieved much higher utilization rates in our restricted problem than is typical in TSN applications; it seems that the additional constraints introduced by the TSN standard are the main culprit behind the difficulty of scheduling. 

Another interesting property is the relationship between the number of tasks and the difficulty of solving the instance. We had problems generating instances that were not quickly solved by the baseline heuristic. For a fixed size of the ``observational interval'', increasing the number of tasks will, at one point, necessarily decrease the average processing time of tasks, which generally makes them easier to schedule.

Finally, we successfully demonstrated how understanding the instances' characteristics can be exploited in this domain. To name a few, using specialized methods for specific underlying support graphs or focusing on optimizing an isolated sub-problem, which was, in our case, a set of tasks corresponding to a single period. 


\section*{Acknowledgments}
This work was supported by the Grant Agency of the Czech Republic under the project GACR 22-31670S and under the project GACR 25-17904S. 

This work was co-funded by the European Union under the project ROBOPROX (reg. no. CZ.02.01.01/00/22\_008/0004590). 

\section*{List of Main Symbols}\label{sec:symbols}

\newcolumntype{L}{>{\raggedright\arraybackslash}X}
\begin{footnotesize}
\setlength{\tabcolsep}{3pt}
\newcolumntype{C}{>{\centering\arraybackslash}X}
\begin{tabularx}{\textwidth}{clcl}
    \textbf{Symbol} & \textbf{Description} &\textbf{Symbol} & \textbf{Description} \\
\hline \endhead
          $\job{k}$ & chain $k$ &
    $\mu$ & number of chains \\
    $\per{\beta}$ & $\beta$-th harmonic period&
    $\hat\hyperp$ & hyper-period \\
    $\larg$ & least period &
    $\perjob{k}$ & period of $\job{k}$\\
    $\jobperset{\tau(\beta)}$ & set of chains with period $\tau(\beta)$ &
    $\task{i}{k}$ & task $i$ of $\job{k}$ \\
    $n^k$ & number of tasks in $\job{k}$ &
    $\proctime{i}{k}$ & processing time of $\task{i}{k}$ \\
    $\pi_i^k$ & assignment of $\task{i}{k}$ to resource &
    $\first{i}{k}$ & start time of first occurrence of $\task{i}{k}$\\
    $\sigma_i^{k}$ & start time of $\task{i}{k}$ in core schedule &
    $\start{i}{k}$ & start time of $\task{i}{k}$ in shifted core schedule\\
    $o_m$ & offset of resource $m$ &
    $\utilization{m}$ & utilization of resource $m$\\
    $\utilinst$ & utilization of the instance &
    $\hyperp$ & height of bin \\
    $\rect{i}{k}$ & rectangle associated with $\task{i}{k}$ &
    $\ell_i^{k}$ & width of $\rect{i}{k}$ \\
    $\height{i}{k}$ & height of $\rect{i}{k}$ &
    $ \orderTasks$ & ordered list of tasks\\
    $\SE{k}$ & start-to-end latency of $\job{k}$ &
    $\DGk{k}$ & degeneracy of $\job{k}$\\
    $\DGkalpha{k}{\alpha}$ & $\alpha$ degeneracy of $\job{k}$ &
    $\DGs$ & degeneracy objective\\
    $\DGm$ & maximum degeneracy objective &
    $\DGsalpha{\alpha}$ & $\alpha$ degeneracy objective \\
\end{tabularx}
\end{footnotesize}

\bibliographystyle{elsarticle-num-names} 
\bibliography{bibperiodic}

\begin{thebibliography}{48}
\expandafter\ifx\csname natexlab\endcsname\relax\def\natexlab#1{#1}\fi
\providecommand{\url}[1]{\texttt{#1}}
\providecommand{\href}[2]{#2}
\providecommand{\path}[1]{#1}
\providecommand{\DOIprefix}{doi:}
\providecommand{\ArXivprefix}{arXiv:}
\providecommand{\URLprefix}{URL: }
\providecommand{\Pubmedprefix}{pmid:}
\providecommand{\doi}[1]{\href{http://dx.doi.org/#1}{\path{#1}}}
\providecommand{\Pubmed}[1]{\href{pmid:#1}{\path{#1}}}
\providecommand{\bibinfo}[2]{#2}
\ifx\xfnm\relax \def\xfnm[#1]{\unskip,\space#1}\fi
\bibitem[{Hanzalek et~al.(2023)Hanzalek, Zahora, and Sojka}]{hanzalek2023}
\bibinfo{author}{Z.~Hanzalek}, \bibinfo{author}{J.~Zahora}, \bibinfo{author}{M.~Sojka},
\newblock \bibinfo{title}{Cone slalom with automated sports car – trajectory planning algorithm},
\newblock \bibinfo{journal}{IEEE Trans. on Vehicular Technology}  (\bibinfo{year}{2023}). \DOIprefix\doi{10.1109/TVT.2023.3309554}.
\bibitem[{Stüber et~al.(2023)Stüber, Osswald, Lindner, and Menth}]{Stuber2023}
\bibinfo{author}{T.~Stüber}, \bibinfo{author}{L.~Osswald}, \bibinfo{author}{S.~Lindner}, \bibinfo{author}{M.~Menth},
\newblock \bibinfo{title}{A survey of scheduling algorithms for the time-aware shaper in time-sensitive networking (tsn)},
\newblock \bibinfo{journal}{IEEE Access} \bibinfo{volume}{11} (\bibinfo{year}{2023}) \bibinfo{pages}{61192--61233}. \DOIprefix\doi{10.1109/ACCESS.2023.3286370}.
\bibitem[{Xue et~al.(2024)Xue, Zhang, Zhou, Nixon, Loveless, and Han}]{Xue2024}
\bibinfo{author}{C.~Xue}, \bibinfo{author}{T.~Zhang}, \bibinfo{author}{Y.~Zhou}, \bibinfo{author}{M.~Nixon}, \bibinfo{author}{A.~Loveless}, \bibinfo{author}{S.~Han},
\newblock \bibinfo{title}{Real-time scheduling for 802.1qbv time-sensitive networking (tsn): A systematic review and experimental study},
\newblock in: \bibinfo{booktitle}{2024 IEEE 30th Real-Time and Embedded Technology and Applications Symposium (RTAS)}, \bibinfo{year}{2024}, pp. \bibinfo{pages}{108--121}. \DOIprefix\doi{10.1109/RTAS61025.2024.00017}.
\bibitem[{Vlk et~al.(2022)Vlk, Brejchová, Hanzálek, and Tang}]{vlk2022largescale}
\bibinfo{author}{M.~Vlk}, \bibinfo{author}{K.~Brejchová}, \bibinfo{author}{Z.~Hanzálek}, \bibinfo{author}{S.~Tang},
\newblock \bibinfo{title}{Large-scale periodic scheduling in time-sensitive networks},
\newblock \bibinfo{journal}{Computers \& Operations Research} \bibinfo{volume}{137} (\bibinfo{year}{2022}) \bibinfo{pages}{105512}. \URLprefix \url{https://www.sciencedirect.com/science/article/pii/S0305054821002549}. \DOIprefix\doi{https://doi.org/10.1016/j.cor.2021.105512}.
\bibitem[{Minaeva and Hanzalek(2021)}]{minaeva2022survey}
\bibinfo{author}{A.~Minaeva}, \bibinfo{author}{Z.~Hanzalek},
\newblock \bibinfo{title}{Survey on periodic scheduling for time-triggered hard real-time systems},
\newblock \bibinfo{journal}{ACM Comput. Surv.} \bibinfo{volume}{54} (\bibinfo{year}{2021}). \URLprefix \url{https://doi.org/10.1145/3431232}. \DOIprefix\doi{10.1145/3431232}.
\bibitem[{Cai and Kong(1996)}]{cai1996nonpreemptive}
\bibinfo{author}{Y.~Cai}, \bibinfo{author}{M.~Kong},
\newblock \bibinfo{title}{Nonpreemptive scheduling of periodic tasks in uni- and multiprocessor systems},
\newblock \bibinfo{journal}{Algorithmica} \bibinfo{volume}{15} (\bibinfo{year}{1996}) \bibinfo{pages}{572--599}.
\bibitem[{Korst et~al.(1996)Korst, Aarts, and Lenstra}]{korst1996scheduling}
\bibinfo{author}{J.~Korst}, \bibinfo{author}{E.~Aarts}, \bibinfo{author}{J.~K. Lenstra},
\newblock \bibinfo{title}{Scheduling periodic tasks},
\newblock \bibinfo{journal}{INFORMS J. Comput.} \bibinfo{volume}{8} (\bibinfo{year}{1996}) \bibinfo{pages}{428--435}.
\bibitem[{Marouf and Sorel(2010)}]{marouf2010schedulability}
\bibinfo{author}{M.~Marouf}, \bibinfo{author}{Y.~Sorel},
\newblock \bibinfo{title}{Schedulability conditions for non-preemptive hard real-time tasks with strict period},
\newblock in: \bibinfo{booktitle}{18th International Conference on Real-Time and Network Systems RTNS'10}, \bibinfo{address}{France, Toulouse}, \bibinfo{year}{2010}, pp. \bibinfo{pages}{50--–58}.
\bibitem[{Marouf and Sorel(2011)}]{marouf2011scheduling}
\bibinfo{author}{M.~Marouf}, \bibinfo{author}{Y.~Sorel},
\newblock \bibinfo{title}{Scheduling non-preemptive hard real-time tasks with strict periods},
\newblock in: \bibinfo{booktitle}{ETFA2011}, \bibinfo{year}{2011}, pp. \bibinfo{pages}{1--8}. \DOIprefix\doi{10.1109/ETFA.2011.6059014}.
\bibitem[{Eisenbrand et~al.(2010)}]{Eisenbrand2010}
\bibinfo{author}{F.~Eisenbrand}, et~al.,
\newblock \bibinfo{title}{Scheduling periodic tasks in a hard real-time environment},
\newblock \bibinfo{journal}{Automata, Languages and Programming} \bibinfo{volume}{6198} (\bibinfo{year}{2010}) \bibinfo{pages}{299--311}.
\bibitem[{Lukasiewycz et~al.(2009)}]{lukasiewycz2009flexray}
\bibinfo{author}{M.~Lukasiewycz}, et~al.,
\newblock \bibinfo{title}{{FlexRay} schedule optimization of the static segment},
\newblock in: \bibinfo{booktitle}{Proc. of IEEE/ACM CODES+ISSS 2009}, \bibinfo{year}{2009}, pp. \bibinfo{pages}{363--372}.
\bibitem[{Hanen and Hanzalek(2020)}]{hanen2020periodic}
\bibinfo{author}{C.~Hanen}, \bibinfo{author}{Z.~Hanzalek},
\newblock \bibinfo{title}{Periodic scheduling and packing problems},
\newblock \bibinfo{journal}{CoRR} \bibinfo{volume}{abs/2011.01898} (\bibinfo{year}{2020}). \URLprefix \url{https://arxiv.org/abs/2011.01898}. \href{http://arxiv.org/abs/2011.01898}{{\tt arXiv:2011.01898}}.
\bibitem[{Grus. et~al.(2024)Grus., Hanen., and Hanzálek.}]{grus2024icores}
\bibinfo{author}{J.~Grus.}, \bibinfo{author}{C.~Hanen.}, \bibinfo{author}{Z.~Hanzálek.},
\newblock \bibinfo{title}{Packing-inspired algorithms for periodic scheduling problems with harmonic periods},
\newblock in: \bibinfo{booktitle}{Proceedings of the 13th International Conference on Operations Research and Enterprise Systems - ICORES}, \bibinfo{organization}{INSTICC}, \bibinfo{publisher}{SciTePress}, \bibinfo{year}{2024}, pp. \bibinfo{pages}{101--112}. \DOIprefix\doi{10.5220/0012325800003639}.
\bibitem[{Eisenbrand et~al.(2010)}]{eisenbrand2010solving}
\bibinfo{author}{F.~Eisenbrand}, et~al.,
\newblock \bibinfo{title}{Solving an avionics real-time scheduling problem by advanced {IP}-methods},
\newblock \bibinfo{journal}{European Symposium on Algorithms}  (\bibinfo{year}{2010}) \bibinfo{pages}{11--22}.
\bibitem[{Al~Sheikh et~al.(2012)Al~Sheikh, Brun, Hladik, and Prabhu}]{AlSheikh2012avionicPeriodic}
\bibinfo{author}{A.~Al~Sheikh}, \bibinfo{author}{O.~Brun}, \bibinfo{author}{P.-E. Hladik}, \bibinfo{author}{B.~J. Prabhu},
\newblock \bibinfo{title}{Strictly periodic scheduling in ima-based architectures},
\newblock \bibinfo{journal}{Real-Time Systems} \bibinfo{volume}{48} (\bibinfo{year}{2012}) \bibinfo{pages}{359--386}. \URLprefix \url{https://doi.org/10.1007/s11241-012-9148-y}. \DOIprefix\doi{10.1007/s11241-012-9148-y}.
\bibitem[{Pira and Artigues(2016)}]{Pira2016gametheorertic}
\bibinfo{author}{C.~Pira}, \bibinfo{author}{C.~Artigues},
\newblock \bibinfo{title}{Line search method for solving a non-preemptive strictly periodic scheduling problem},
\newblock \bibinfo{journal}{Journal of Scheduling} \bibinfo{volume}{19} (\bibinfo{year}{2016}) \bibinfo{pages}{227--243}. \URLprefix \url{https://doi.org/10.1007/s10951-014-0389-6}. \DOIprefix\doi{10.1007/s10951-014-0389-6}.
\bibitem[{Deroche et~al.(2017)Deroche, Scharbarg, and Fraboul}]{deroche2017avionicGreedyWithEnumearation}
\bibinfo{author}{E.~Deroche}, \bibinfo{author}{J.-L. Scharbarg}, \bibinfo{author}{C.~Fraboul},
\newblock \bibinfo{title}{A greedy heuristic for distributing hard real-time applications on an ima architecture},
\newblock in: \bibinfo{booktitle}{2017 12th IEEE International Symposium on Industrial Embedded Systems (SIES)}, \bibinfo{year}{2017}, pp. \bibinfo{pages}{1--8}. \DOIprefix\doi{10.1109/SIES.2017.7993390}.
\bibitem[{Blikstad et~al.(2018)Blikstad, Karlsson, L{\"o}{\"o}w, and R{\"o}nnberg}]{Blikstad2018relaxedsubaero}
\bibinfo{author}{M.~Blikstad}, \bibinfo{author}{E.~Karlsson}, \bibinfo{author}{T.~L{\"o}{\"o}w}, \bibinfo{author}{E.~R{\"o}nnberg},
\newblock \bibinfo{title}{An optimisation approach for pre-runtime scheduling of tasks and communication in an integrated modular avionic system},
\newblock \bibinfo{journal}{Optimization and Engineering} \bibinfo{volume}{19} (\bibinfo{year}{2018}) \bibinfo{pages}{977--1004}. \URLprefix \url{https://doi.org/10.1007/s11081-018-9385-6}. \DOIprefix\doi{10.1007/s11081-018-9385-6}.
\bibitem[{Monot et~al.(2012)Monot, Navet, Bavoux, and Simonot-Lion}]{monot2012autosarLptAndPartitioning}
\bibinfo{author}{A.~Monot}, \bibinfo{author}{N.~Navet}, \bibinfo{author}{B.~Bavoux}, \bibinfo{author}{F.~Simonot-Lion},
\newblock \bibinfo{title}{Multisource software on multicore automotive ecus—combining runnable sequencing with task scheduling},
\newblock \bibinfo{journal}{IEEE Transactions on Industrial Electronics} \bibinfo{volume}{59} (\bibinfo{year}{2012}) \bibinfo{pages}{3934--3942}. \DOIprefix\doi{10.1109/TIE.2012.2185913}.
\bibitem[{Becker et~al.(2016)Becker, Dasari, Nicolic, Akesson, Nélis, and Nolte}]{becker2016autosarIlp}
\bibinfo{author}{M.~Becker}, \bibinfo{author}{D.~Dasari}, \bibinfo{author}{B.~Nicolic}, \bibinfo{author}{B.~Akesson}, \bibinfo{author}{V.~Nélis}, \bibinfo{author}{T.~Nolte},
\newblock \bibinfo{title}{Contention-free execution of automotive applications on a clustered many-core platform},
\newblock in: \bibinfo{booktitle}{2016 28th Euromicro Conference on Real-Time Systems (ECRTS)}, \bibinfo{year}{2016}, pp. \bibinfo{pages}{14--24}. \DOIprefix\doi{10.1109/ECRTS.2016.14}.
\bibitem[{Dvorak and Hanzalek(2014)}]{dvorak2014multi}
\bibinfo{author}{J.~Dvorak}, \bibinfo{author}{Z.~Hanzalek},
\newblock \bibinfo{title}{Multi-variant time constrained {FlexRay} static segment scheduling},
\newblock \bibinfo{journal}{2014 10th IEEE WFCS}  (\bibinfo{year}{2014}) \bibinfo{pages}{1--8}.
\bibitem[{Dvořák and Hanzálek(2019)}]{dvorak2019}
\bibinfo{author}{J.~Dvořák}, \bibinfo{author}{Z.~Hanzálek},
\newblock \bibinfo{title}{Multi-variant scheduling of critical time-triggered communication in incremental development process: Application to flexray},
\newblock \bibinfo{journal}{IEEE Transactions on Vehicular Technology} \bibinfo{volume}{68} (\bibinfo{year}{2019}) \bibinfo{pages}{155--169}. \DOIprefix\doi{10.1109/TVT.2018.2879920}.
\bibitem[{Liao and Chen(2003)}]{Liao2003CorPerMaintanenca}
\bibinfo{author}{C.~Liao}, \bibinfo{author}{W.~Chen},
\newblock \bibinfo{title}{Single-machine scheduling with periodic maintenance and nonresumable jobs},
\newblock \bibinfo{journal}{Computers \& Operations Research} \bibinfo{volume}{30} (\bibinfo{year}{2003}) \bibinfo{pages}{1335--1347}. \URLprefix \url{https://www.sciencedirect.com/science/article/pii/S0305054802000746}. \DOIprefix\doi{https://doi.org/10.1016/S0305-0548(02)00074-6}.
\bibitem[{Xu et~al.(2008)Xu, Sun, and Li}]{xu2008parallelmaint}
\bibinfo{author}{D.~Xu}, \bibinfo{author}{K.~Sun}, \bibinfo{author}{H.~Li},
\newblock \bibinfo{title}{Parallel machine scheduling with almost periodic maintenance and non-preemptive jobs to minimize makespan},
\newblock \bibinfo{journal}{Computers\& Operations Research} \bibinfo{volume}{35} (\bibinfo{year}{2008}) \bibinfo{pages}{1344--1349}. \URLprefix \url{https://www.sciencedirect.com/science/article/pii/S0305054806002036}. \DOIprefix\doi{https://doi.org/10.1016/j.cor.2006.08.015}.
\bibitem[{Wu et~al.(2014)Wu, Dong, and Zheng}]{wu2014patient}
\bibinfo{author}{Y.~Wu}, \bibinfo{author}{M.~Dong}, \bibinfo{author}{Z.~Zheng},
\newblock \bibinfo{title}{Patient scheduling with periodic deteriorating maintenance on single medical device},
\newblock \bibinfo{journal}{Computers\& Operations Research} \bibinfo{volume}{49} (\bibinfo{year}{2014}) \bibinfo{pages}{107--116}. \URLprefix \url{https://www.sciencedirect.com/science/article/pii/S0305054814000884}. \DOIprefix\doi{https://doi.org/10.1016/j.cor.2014.04.005}.
\bibitem[{Roy et~al.(2016)Roy, Zhang, Chang, Goswami, and Chakraborty}]{roy2016multi}
\bibinfo{author}{D.~Roy}, \bibinfo{author}{L.~Zhang}, \bibinfo{author}{W.~Chang}, \bibinfo{author}{D.~Goswami}, \bibinfo{author}{S.~Chakraborty},
\newblock \bibinfo{title}{Multi-objective co-optimization of {FlexRay}-based distributed control systems},
\newblock in: \bibinfo{booktitle}{2016 IEEE Real-Time and Embedded Technology and Applications Symposium (RTAS)}, \bibinfo{publisher}{IEEE}, \bibinfo{address}{Austria, Vienna}, \bibinfo{year}{2016}, pp. \bibinfo{pages}{1--12}.
\bibitem[{Minaeva et~al.(2021)Minaeva, Roy, Akesson, Hanzálek, and Chakraborty}]{minaeva2021control}
\bibinfo{author}{A.~Minaeva}, \bibinfo{author}{D.~Roy}, \bibinfo{author}{B.~Akesson}, \bibinfo{author}{Z.~Hanzálek}, \bibinfo{author}{S.~Chakraborty},
\newblock \bibinfo{title}{Control performance optimization for application integration on automotive architectures},
\newblock \bibinfo{journal}{IEEE Transactions on Computers} \bibinfo{volume}{70} (\bibinfo{year}{2021}) \bibinfo{pages}{1059--1073}. \DOIprefix\doi{10.1109/TC.2020.3003083}.
\bibitem[{Barth et~al.(2018)Barth, Guiraud, Leclerc, Marcé, and Strozecki}]{guiraud2018}
\bibinfo{author}{D.~Barth}, \bibinfo{author}{M.~Guiraud}, \bibinfo{author}{B.~Leclerc}, \bibinfo{author}{O.~Marcé}, \bibinfo{author}{Y.~Strozecki},
\newblock \bibinfo{title}{Deterministic scheduling of periodic messages for cloud ran},
\newblock in: \bibinfo{booktitle}{2018 25th International Conference on Telecommunications (ICT)}, \bibinfo{year}{2018}, pp. \bibinfo{pages}{405--410}. \DOIprefix\doi{10.1109/ICT.2018.8464890}.
\bibitem[{Guiraud and Strozecki(2024)}]{GuiraudS24}
\bibinfo{author}{M.~Guiraud}, \bibinfo{author}{Y.~Strozecki},
\newblock \bibinfo{title}{Scheduling periodic messages on a shared link without buffering},
\newblock \bibinfo{journal}{J. Sched.} \bibinfo{volume}{27} (\bibinfo{year}{2024}) \bibinfo{pages}{461--484}. \URLprefix \url{https://doi.org/10.1007/s10951-024-00813-0}. \DOIprefix\doi{10.1007/S10951-024-00813-0}.
\bibitem[{Steiner(2008)}]{steiner2008ttethernet}
\bibinfo{author}{W.~Steiner},
\newblock \bibinfo{title}{{TTE}thernet specification},
\newblock \bibinfo{journal}{TTTech Computertechnik AG, Nov} \bibinfo{volume}{39} (\bibinfo{year}{2008}) \bibinfo{pages}{40}.
\bibitem[{Serna~Oliver et~al.(2018)Serna~Oliver, Craciunas, and Steiner}]{serna2018tsnwindows}
\bibinfo{author}{R.~Serna~Oliver}, \bibinfo{author}{S.~S. Craciunas}, \bibinfo{author}{W.~Steiner},
\newblock \bibinfo{title}{Ieee 802.1qbv gate control list synthesis using array theory encoding},
\newblock in: \bibinfo{booktitle}{2018 IEEE Real-Time and Embedded Technology and Applications Symposium (RTAS)}, \bibinfo{year}{2018}, pp. \bibinfo{pages}{13--24}. \DOIprefix\doi{10.1109/RTAS.2018.00008}.
\bibitem[{Hellmanns et~al.(2020)Hellmanns, Glavackij, Falk, Hummen, Kehrer, and Dürr}]{hellmanns2020industrial}
\bibinfo{author}{D.~Hellmanns}, \bibinfo{author}{A.~Glavackij}, \bibinfo{author}{J.~Falk}, \bibinfo{author}{R.~Hummen}, \bibinfo{author}{S.~Kehrer}, \bibinfo{author}{F.~Dürr},
\newblock \bibinfo{title}{Scaling tsn scheduling for factory automation networks},
\newblock in: \bibinfo{booktitle}{2020 16th IEEE International Conference on Factory Communication Systems (WFCS)}, \bibinfo{year}{2020}, pp. \bibinfo{pages}{1--8}. \DOIprefix\doi{10.1109/WFCS47810.2020.9114415}.
\bibitem[{Reusch et~al.(2022)Reusch, Craciunas, and Pop}]{reusch2022tsnwithauth}
\bibinfo{author}{N.~Reusch}, \bibinfo{author}{S.~S. Craciunas}, \bibinfo{author}{P.~Pop},
\newblock \bibinfo{title}{Dependability-aware routing and scheduling for time-sensitive networking},
\newblock \bibinfo{journal}{IET Cyber-Physical Systems: Theory \& Applications} \bibinfo{volume}{7} (\bibinfo{year}{2022}) \bibinfo{pages}{124--146}. \DOIprefix\doi{https://doi.org/10.1049/cps2.12030}.
\bibitem[{Vlk et~al.(2020)Vlk, Hanz{\'a}lek, Brejchov{\'a}, Tang, Bhattacharjee, and Fu}]{vlk2020enhancing}
\bibinfo{author}{M.~Vlk}, \bibinfo{author}{Z.~Hanz{\'a}lek}, \bibinfo{author}{K.~Brejchov{\'a}}, \bibinfo{author}{S.~Tang}, \bibinfo{author}{S.~Bhattacharjee}, \bibinfo{author}{S.~Fu},
\newblock \bibinfo{title}{Enhancing schedulability and throughput of time-triggered traffic in {IEEE 802.1 Qbv} time-sensitive networks},
\newblock \bibinfo{journal}{IEEE Transactions on Communications}  (\bibinfo{year}{2020}).
\bibitem[{Hlad{\'i}k et~al.(2020)Hlad{\'i}k, Minaeva, and Hanz{\'a}lek}]{hladik2020complexity}
\bibinfo{author}{R.~Hlad{\'i}k}, \bibinfo{author}{A.~Minaeva}, \bibinfo{author}{Z.~Hanz{\'a}lek},
\newblock \bibinfo{title}{On the complexity of a periodic scheduling problem with precedence relations},
\newblock in: \bibinfo{booktitle}{Combinatorial Optimization and Applications}, \bibinfo{year}{2020}, pp. \bibinfo{pages}{107--124}.
\bibitem[{Grus et~al.(2025)Grus, Hanzalek, and Hanen}]{our-data}
\bibinfo{author}{J.~Grus}, \bibinfo{author}{Z.~Hanzalek}, \bibinfo{author}{C.~Hanen}, \bibinfo{title}{Periodic chains scheduling on dedicated resources - a crucial problem in time-sensitive networks - supplemental material}, \bibinfo{year}{2025}. \URLprefix \url{https://doi.org/10.5281/zenodo.14901765}. \DOIprefix\doi{10.5281/zenodo.14901765}.
\bibitem[{Guiraud(2021)}]{guiraudthese}
\bibinfo{author}{M.~Guiraud}, \bibinfo{title}{{Deterministic scheduling of periodic datagrams for low latency in 5G and beyond}}, \bibinfo{type}{Theses}, {Universit{\'e} Paris-Saclay}, \bibinfo{year}{2021}. \URLprefix \url{https://theses.hal.science/tel-03413419}.
\bibitem[{Hanen and Hanzalek(2012)}]{hanen2012grouping}
\bibinfo{author}{C.~Hanen}, \bibinfo{author}{Z.~Hanzalek},
\newblock \bibinfo{title}{Grouping tasks to save energy in a cyclic scheduling problem: a complexity study},
\newblock \bibinfo{year}{2012}.
\bibitem[{Ogata(1987)}]{ogata-1987a}
\bibinfo{author}{K.~Ogata}, \bibinfo{title}{Discrete-Time Control Systems}, \bibinfo{publisher}{Prentice Hall}, \bibinfo{address}{Australia, Sydney}, \bibinfo{year}{1987}.
\bibitem[{Grama et~al.(2003)Grama, Karypis, Kumar, and Gupta}]{grama03}
\bibinfo{author}{A.~Grama}, \bibinfo{author}{G.~Karypis}, \bibinfo{author}{V.~Kumar}, \bibinfo{author}{A.~Gupta}, \bibinfo{title}{Introduction to Parallel Computing}, \bibinfo{edition}{second} ed., \bibinfo{publisher}{Addison-Wesley}, \bibinfo{year}{2003}.
\bibitem[{Johnson(1954)}]{Johnson54}
\bibinfo{author}{S.~M. Johnson},
\newblock \bibinfo{title}{Optimal two- and three-stage production schedules with setup times included},
\newblock \bibinfo{journal}{Naval Research Logistics Quarterly} \bibinfo{volume}{1} (\bibinfo{year}{1954}) \bibinfo{pages}{61--68}. \DOIprefix\doi{https://doi.org/10.1002/nav.3800010110}.
\bibitem[{Reddi and Ramamoorthy(1972)}]{Reddi01091972}
\bibinfo{author}{S.~S. Reddi}, \bibinfo{author}{C.~V. Ramamoorthy},
\newblock \bibinfo{title}{On the flow-shop sequencing problem with no wait in process†},
\newblock \bibinfo{journal}{Journal of the Operational Research Society} \bibinfo{volume}{23} (\bibinfo{year}{1972}) \bibinfo{pages}{323--331}. \URLprefix \url{https://doi.org/10.1057/jors.1972.52}. \DOIprefix\doi{10.1057/jors.1972.52}. \href{http://arxiv.org/abs/https://doi.org/10.1057/jors.1972.52}{{\tt arXiv:https://doi.org/10.1057/jors.1972.52}}.
\bibitem[{Gilmore and Gomory(1964)}]{gilmore1964sequencing}
\bibinfo{author}{P.~C. Gilmore}, \bibinfo{author}{R.~E. Gomory},
\newblock \bibinfo{title}{Sequencing a one state-variable machine: A solvable case of the traveling salesman problem},
\newblock \bibinfo{journal}{Operations research} \bibinfo{volume}{12} (\bibinfo{year}{1964}) \bibinfo{pages}{655--679}.
\bibitem[{Hall and Sriskandarajah(1996)}]{HallS96}
\bibinfo{author}{N.~G. Hall}, \bibinfo{author}{C.~Sriskandarajah},
\newblock \bibinfo{title}{A survey of machine scheduling problems with blocking and no-wait in process},
\newblock \bibinfo{journal}{Oper. Res.} \bibinfo{volume}{44} (\bibinfo{year}{1996}) \bibinfo{pages}{510--525}. \URLprefix \url{https://doi.org/10.1287/opre.44.3.510}. \DOIprefix\doi{10.1287/OPRE.44.3.510}.
\bibitem[{Shaw(2004)}]{Shaw2004cpconstraint}
\bibinfo{author}{P.~Shaw},
\newblock \bibinfo{title}{A constraint for bin packing},
\newblock in: \bibinfo{booktitle}{International Conference on Principles and Practice of Constraint Programming}, \bibinfo{year}{2004}. \URLprefix \url{https://api.semanticscholar.org/CorpusID:11225063}.
\bibitem[{Della Croce et~al.(2014)Della Croce, Grosso, and Salassa}]{della-matheur}
\bibinfo{author}{F.~Della Croce}, \bibinfo{author}{A.~Grosso}, \bibinfo{author}{F.~Salassa},
\newblock \bibinfo{title}{A matheuristic approach for the two-machine total completion time flow shop problem},
\newblock \bibinfo{journal}{Annals of Operations Research} \bibinfo{volume}{213} (\bibinfo{year}{2014}) \bibinfo{pages}{67--78}. \URLprefix \url{https://doi.org/10.1007/s10479-011-0928-x}. \DOIprefix\doi{10.1007/s10479-011-0928-x}.
\bibitem[{Christiaens and Vanden~Berghe(2020)}]{greet2020RR}
\bibinfo{author}{J.~Christiaens}, \bibinfo{author}{G.~Vanden~Berghe},
\newblock \bibinfo{title}{Slack induction by string removals for vehicle routing problems},
\newblock \bibinfo{journal}{Transportation Science} \bibinfo{volume}{54} (\bibinfo{year}{2020}). \DOIprefix\doi{10.1287/trsc.2019.0914}.
\bibitem[{Lo~Bello et~al.(2023)Lo~Bello, Patti, and Leonardi}]{zonal}
\bibinfo{author}{L.~Lo~Bello}, \bibinfo{author}{G.~Patti}, \bibinfo{author}{L.~Leonardi},
\newblock \bibinfo{title}{A perspective on ethernet in automotive communications—current status and future trends},
\newblock \bibinfo{journal}{Applied Sciences} \bibinfo{volume}{13} (\bibinfo{year}{2023}) \bibinfo{pages}{1278}. \DOIprefix\doi{10.3390/app13031278}.

\end{thebibliography}
\end{document}